\documentclass[%
 preprint,%
 amssymb, amsmath,%
 aip,jcp,%cha,%
]{revtex4-1}
\usepackage{graphicx}
\usepackage{inputenc}
\usepackage{bm}%
\usepackage{longtable}
\usepackage{supertabular}
\usepackage{rotating}
\usepackage{ulem} % DL
\usepackage[colorlinks=false,linkcolor=blue]{hyperref}%    set false by aph
\usepackage{wrapfig}
\usepackage{ifthen}
\newboolean{includeMQDT}
\setboolean{includeMQDT}{false}  %% edit this line to have either "true" or "false" in the brackets
\usepackage[dvipsnames]{xcolor}

\usepackage{dcolumn,ulem}  %  added by aph
\usepackage{xspace}

\begin{document}
\bibliographystyle{jcp}

\title{\large HF trimer: 12D fully coupled quantum calculations of HF-stretch excited intramolecular and intermolecular vibrational states using contracted bases of intramolecular and intermolecular eigenstates}
\author{Peter M. Felker}
\email{felker@chem.ucla.edu}
\affiliation{Department of Chemistry and Biochemistry, University of California, Los Angeles, CA 90095-1569, USA}
\author{Zlatko Ba\v{c}i\'{c}}
\email{zlatko.bacic@nyu.edu}
\affiliation{Department of Chemistry, New York University, New York, New York 10003, USA}
\affiliation{Simons Center for Computational Physical Chemistry at New York University}
\affiliation{NYU-ECNU Center for Computational Chemistry at NYU Shanghai, 3663 Zhongshan Road North, Shanghai, 200062, China}
\date{\today}

\begin{abstract}

We present the computational methodology which for the first time allows rigorous twelve-dimensional (12D) quantum calculations of the coupled intramolecular and intermolecular vibrational states of hydrogen-bonded trimers of flexible diatomic molecules. Its starting point is the approach that we introduced recently for fully coupled 9D quantum calculations of the intermolecular vibrational states of noncovalently bound trimers comprised of diatomics treated as rigid. In this paper it is extended to include the intramolecular stretching coordinates of the three diatomic monomers.  The cornerstone of our 12D methodology is the partitioning of the full vibrational Hamiltonian of the trimer into two reduced-dimension Hamiltonians, one in 9D for the intermolecular degrees of freedom (DOFs)  and another in 3D for the intramolecular vibrations of the trimer, and a remainder term. These two Hamiltonians are diagonalized separately and a fraction of their respective 9D and 3D eigenstates is included in the 12D product contracted basis for both the intra- and intermolecular DOFs, in which the matrix of the full 12D vibrational Hamiltonian of the trimer is diagonalized. This methodology is implemented in the 12D quantum calculations of the coupled intra- and intermolecular vibrational states of the hydrogen-bonded HF trimer on an {\it ab initio} calculated potential energy surface (PES). The calculations encompass the one- and two-quanta intramolecular HF-stretch excited vibrational states of the trimer and low-energy intermolecular vibrational states in the intramolecular vibrational manifolds of interest. They reveal several interesting manifestations of significant coupling between the intra- and intermolecular vibrational modes of (HF)$_3$. The 12D calculations also show that the frequencies of the $v=1,2$ HF stretching states of HF trimer are strongly redshifted in comparison to those of the isolated HF monomer. Moreover, the magnitudes of these trimer redshifts are much larger than that of the redshift for the stretching fundamental of the donor-HF moiety in (HF)$_2$, most likely due to the cooperative hydrogen bonding in (HF)$_3$. The agreement between the 12D results and the limited spectroscopic data for HF trimer, while satisfactory, leaves room for improvement and points to the need for a more accurate PES.

\end{abstract}
\maketitle

\section{Introduction}\label{sec_introduction}

Noncovalently bound molecular complexes resulting from hydrogen bonding or Van der Waals interactions between the constituent monomers have attracted the attention of experimentalists and theorists for decades. To date, high-resolution spectroscopic studies and theoretical treatments of the rovibrational states and infrared and Raman spectra have largely been focused on the weakly bound binary molecular complexes. In a large majority of quantum bound-state calculations of these systems the monomers have been treated as rigid, taking advantage of the fact that the frequencies of the intramolecular monomer vibrations  are generally much higher than those of the intermolecular vibrations of the complexes.\cite{BACIC96C,AVOIRD00A,CARRING11A,AVOIRD22,MATYUS23} The rigid-monomer approximation reduces significantly the dimensionality and the cost of the computations relative to that for flexible monomers. Moreover, accurate potential energy surfaces (PESs) available for weakly bound complexes where the monomers are taken to be rigid greatly outnumber full-dimensional dimer PESs obtained for flexible monomers. However, the rigid-monomer calculations suffer from a number of fundamental limitations: (1) The neglect of coupling between the intramolecular and intermolecular degrees of freedom (DOFs) inevitably introduces errors of hard-to-determine magnitude in all the results (bound states, tunneling splittings) obtained in this way already for monomers in their ground states. (2) It is impossible to calculate intramolecular vibrational frequencies and their complexation-induced shifts from the gas-phase monomer values, as well as the effects of intramolecular vibrational excitations on the intermolecular vibrational frequencies and tunneling splittings. These important and often measured spectroscopic properties can be obtained only from rigorous quantum calculations in full dimensionality, where the monomers are treated as flexible. 

The first steps in this direction were taken already a couple of decades ago, with the fully coupled quantum six-dimensional (6D) calculations of the (ro)vibrational states of (HF)$_2$, (DF)$_2$, and HFDF,\cite{BACIC95,BACIC98A} as well as (HCl)$_2$,\cite{BACIC97A,BACIC98A} for the monomers in their ground vibrational states. But extending such full-dimensional calculations to the case when one or both monomers are vibrationally excited presented an entirely new set of challenges. One of them is a greatly increased dimensionality of the problem, since accurate description of  vibrationally excited monomers in the dimer requires a basis for the intramolecular DOFs that is considerably larger than when they are in the ground state. The second challenge stems from the order-of-magnitude (or greater) difference between the high frequencies of the intramolecular vibrations and those (much lower) of the intermolecular vibrations typical for noncovalently bound molecular complexes. This gives rise to a high density of intermolecular vibrational states below and at the energies of the intramolecular fundamentals and overtones, which was thought for a long time to present a daunting obstacle to full-dimensional quantum bound-state calculations  for excited monomer vibrations. As a result, until a couple of years ago only a few such calculations were reported, for HF-stretch excited (HF)$_2$\cite{WU95,AVOIRD03} and the HCl-stretch excited (HCl)$_2$.\cite{BACIC981} With greatly increased computing power, it became possible, at great computational effort, to compute directly, from the ground state up, the 6D monomer-excited states of (HF)$_2$,\cite{HUANG19} and also the vibrational states of (H$_2$O)$_2$ with monomer-bend excitation in 12D.\cite{CARRING18} For the latter, the density of states in the region of the energies of the monomer stretch excitations was prohibitively high for the computational approach employed, which used contracted basis functions for the intramolecular vibrational DOFs but not for the intermolecular coordinates.\cite{CARRING18} 

The situation changed with the introduction of the general approach for full-dimensional and fully coupled quantum computation of (ro)vibrational states of noncovalently bound binary molecular complexes,\cite{BACIC19B,BACIC22C} in which contracted basis functions are used for {\it both} intramolecular and intermolecular DOFs. In this computational scheme, the full rovibrational Hamiltonian of the binary molecular complex is partitioned into a rigid-monomer intermolecular vibrational Hamiltonian, two intramolecular vibrational Hamiltonians -- one for each monomer, and a  remainder term. Each of the three reduced-dimension Hamiltonians is diagonalized separately. Only a small fraction of the intermolecular eigenstates, with energies much lower than those of the intramolecular vibrational excitations of interest, is included, together with the selected intramolecular eigenstates, in a compact final product contracted basis covering all internal, intra- and intermolecular, DOFs of the complex. The use of contracted eigenstate bases for intramolecular and intermolecular coordinates makes it particularly easy to take advantage of the key insight regarding the extremely weak coupling between the two sets of internal DOFs, that emerged from the quantum 6D calculations  of the vibration-translation-rotation eigenstates of flexible H$_2$, HD, and D$_2$ inside the clathrate hydrate cage.\cite{BACIC19A} This methodology, initially applied to the 6D monomer-excited vibrational states of (HF)$_2$,\cite{BACIC19B} was soon extended to the 9D rovibrational states of the triatom-diatom noncovalently bound complexes.\cite{BACIC20C} Its effectiveness was demonstrated by the fully coupled 9D quantum calculations of the $J=0,1$  rovibrational states of H$_2$O-CO and D$_2$O-CO complexes, including those associated with all monomer intramolecular vibrational fundamentals,\cite{BACIC20C} the first of this kind for a noncovalently bound molecular complex with more than four atoms.

The eigenstates of intermediate reduced-dimension Hamiltonians have long been used in various computational schemes devised to decrease the size of the final full-dimensional basis for bound-state calculations. The sequential diagonalization-truncation method of Ba\v{c}i\'{c}  and Light\cite{BACIC86,BACIC87,BACIC88,BACIC89} was applied very successfully to floppy isomerizing molecules, e.g., LiCN/LiNC and HCN/HNC, as well as weakly bound molecular complexes. Carrington and co-workers\cite{CARRING02,CARRING03,CARRING06,CARRING08} have, on the other hand, developed approaches in which the internal coordinates of (covalently bound) polyatomic molecules are divided in two groups, referred to as stretch and bend, and contracted basis functions are used for both groups, subject to two different energy cutoffs. However, they did not extend this approach to noncovalently bound molecular complexes. Zou, Bowman, and Brown,\cite{BOWMAN03A} in their quantum 6D calculations of the vibrational states of the isomerizing acetylene/vinylidene system to energies above the threshold for the formation of vinylidene states, combined the eigenfunctions of the physically motivated reduced-dimension 4D and 2D Hamiltonians into the final 6D basis in which the full-dimensional vibrational Hamiltonian was diagonalized. A certain number of the intermediate 4D eigenstates below and above the lowest-energy (4D) vinylidene state was included in the 6D basis. 

In a rather short time, the methodology of Felker and Ba\v{c}i\'{c}\cite{BACIC19B,BACIC20C,BACIC22C} described above, that uses contracted eigenstate bases for both intermolecular  and intramolecular coordinates, has proved to be remarkably versatile.\cite{BACIC22C} Besides H$_2$O/D$_2$O-CO,\cite{BACIC20C} it has enabled full-dimensional (9D) and fully coupled quantum treatments of other noncovalently bound triatom-diatom complexes in monomer-excited states, such as HDO-CO,\cite{BACIC21} H$_2$O-HCl\cite{BACIC21A} and several of its H/D isotopologues.\cite{BACIC21B,BACIC21C} In addition, this approach was implemented successfully in rigorous 8D quantum calculations of the intramolecular stretch fundamentals of two H$_2$ molecules in the large clathrate hydrate cage,\cite{BACIC19D} the fully coupled 9D quantum treatment of the intra- and intermolecular vibrational levels of flexible H$_2$O in (rigid) C$_{60}$,\cite{BACIC20} and the fully coupled 9D quantum calculations of flexible H$_2$O/HDO intramolecular excitations and intermolecular states of the benzene-H$_2$O and benzene-HDO complexes (for rigid benzene).\cite{BACIC20A} Finally, very recently, Wang and Carrington have employed contracted intermolecular and intramolecular basis functions, comprised of the eigenstates of the corresponding reduced-dimension Hamiltonians, to calculate in 12D the excited OH-stretch states of the water dimer.\cite{CARRING23}

All this naturally raises the question whether these and similar approaches can be extended to noncovalently bound molecular trimers and, eventually, larger molecular clusters. Among noncovalently bound molecular clusters, molecular trimers are of particular interest, as the smallest clusters in which the nonadditive many-body (three-body in this case) interactions can manifest. Their accurate description is of great importance, since they play a crucial role in shaping the structural and dynamical properties of condensed phases. Molecular trimers provide the best opportunity for testing the accuracy of the computed three-body interactions through comparison of high-level quantum bound-state calculations on the potential surfaces (PESs) that incorporate them with the spectroscopic data. But the methodology that existed until recently was rudimentary and not up to the task. Prior to our study described below,\cite{BACIC22B} the only molecular trimer for which fully coupled quantum bound-state calculations were reported about 20 years ago was the very weakly bound (H$_2$)$_3$, in the rigid-monomers approximation\cite{CLARY02} and in full dimensionality.\cite{YU04}  Both calculations found only one bound state for each symmetry, and the computational approaches employed were not applied to any other more strongly bound molecular trimer.

Clearly, it is desirable to develop rigorous bound-state methodologies with a wider range of applicability, capable of treating accurately more strongly bound, hydrogen-bonded (HB) molecular trimers with many excited intra- and intermolecular vibrational states. HB trimers formed by diatomic molecules, e.g., (HF)$_3$ and (HCl)$_3$, present a more realistic and tractable initial target than those involving triatomic molecules, such as (H$_2$O)$_3$. This is primarily due to the lower dimensionality of the former, 12D for flexible monomers and 9D if they are taken to be rigid. In comparison, for (H$_2$O)$_3$,  the problem is 21D for flexible monomers, and 12D in the rigid-monomer approximation. But even for HF trimer, the two variational bound-state calculations in the literature treated only the intermolecular bending (or torsional) levels, while the remaining DOFs, the three intermonomer center-of-mass (c.m.) distances and HF bond lengths were held fixed. The first such study by Kolebrander {\it et al.}\cite{LISY88} treated the three in-plane bends as decoupled from the three out-of plane bends, leading to two separate 3D quantum calculations. Later, Wang and Carrington went one step further and performed 6D quantum calculations of the bending levels of (HF)$_3$ and (DF)$_3$,\cite{CARRING01} treating the in-plane and out-of-plane bends as coupled, and all other degrees of freedom (DOFs), intra- and intermolecular, as frozen. Consequently, left unanswered by both studies was the key question of the extent of the coupling between the intermolecular bending and stretching DOFs of the trimer, and its effects on the intermolecular vibrational states.

In order to fill this methodological gap very recently in Ref. \onlinecite{BACIC22B}, hereafter denoted as I, we introduced a computational approach allowing efficient and rigorous 9D quantum calculations of the intermolecular vibrational states of general noncovalently bound trimers of diatomic molecules. The more strongly bound molecular trimers, such as the hydrogen-bonded (HF)$_3$ and (HCl)$_3$, where three-body interactions are expected to be important, are the main intended targets of this approach. The intermolecular trimer coordinates and the corresponding rigid-monomer 9D vibrational $(J=0)$  Hamiltonian of Wang and Carrington\cite{CARRING01B} are employed. In this treatment, the intermolecular stretching and bending vibrations are fully coupled, and the only dynamical approximation made is to keep the bond lengths of the three monomers fixed.\cite{BACIC22B}  

In I, we chose to implement the new methodology first on (HF)$_3$, for several reasons. Chief among them were its stature as a paradigmatic hydrogen-bonded trimer and the availability of a full-dimensional PES. From several PESs constructed by Quack, Stohner, and Suhm for (HF)$_3$ and larger HF aggregates,\cite{QUACK93,QUACK01,QUACK98A} all of them represented as a sum of many-body terms, we selected the one that combines the SO-3 two-body potential\cite{QUACK98} with the three-body term designated HF3BG.\cite{QUACK01} Wang and Carrington used the same PES in their quantum 6D calculations of the bending energy eigenstates of (HF)$_3$,\cite{CARRING01} so that direct comparison was possible between their 6D results and ours in 9D. On the experimental side, the infrared (IR) spectra of the HF and HD trimers in supersonic molecular beams demonstrated that the two isotopologues have a cyclic structure which, due to the vibrational averaging, is that of an oblate symmetric top.\cite{NESBITT93,ROYP14} IR spectroscopy of the trimers in molecular beams also led to the determination of their degenerate HF and DF stretch fundamentals,\cite{LISY86,LISY88,NESBITT93} and revealed two bands attributed to the in-plane and out-of plane bending fundamentals, respectively.\cite{ROYP14} Additional limited but valuable experimental information regarding the intermolecular vibrations of the HF and DF trimers comes from the IR spectra in Ar and Ne matrices.\cite{ANDREWS84,ANDREWS92,ANDREWS99} 

The quantum 9D rigid-monomer calculations on the HF trimer in I yielded frequencies of the intermolecular bending fundamentals that are about 10\% lower than those those from the earlier 6D quantum calculations that considered only the bending modes of the trimer.\cite{CARRING01} This demonstrated conclusively that the stretch-bend coupling in (HF)$_3$ is strong and must be included in any quantitative treatment of its vibrations, assessment of the quality of the PES employed, and comparison with spectroscopic results. Moreover, the 9D quantum calculations performed on the 2-body (SO-3) trimer PES, in which the 3-body term is not included, gave vibrational energies which are invariably significantly below those obtained for the 2+3-body SO-3 + HF3BG PES (and also in worse agreement with the available experimental data). This led to the conclusion that the contribution which the 3-body interaction makes to the (HF)$_3$ PES is large, and its inclusion is mandatory if accurate results are desired.

However, only rigorous full-dimensional quantum calculations of their vibrational levels can provide definitive results for the PES employed and unambiguous assessment of the quality of the PES by comparison with the relevant spectroscopic data. Therefore, here we take this final step and, extending the rigid-monomer methodology in I, present the computational approach for full-dimensional (12D) and fully coupled quantum calculations of the inter- and intramolecular vibrational states of noncovalently bound trimers of flexible diatomic molecules. It incorporates the key element on the methodology introduced by us\cite{BACIC19B,BACIC20C} and implemented successfully on binary molecular complexes\cite{BACIC22C} - the use of contracted basis functions for both intermolecular and intramolecular DOFs of the trimer. These are obtained by diagonalizing separately a 9D rigid-monomer intermolecular vibrational Hamiltonian (as done in I) and a 3D intramolecular vibrational Hamiltonian for the three monomers, respectively. A fraction of the lower-energy eigenstates of these two reduced-dimension Hamiltonians is included in the final 12D product contracted basis covering all intra- and intermolecular DOFs in which the matrix of the full vibrational Hamiltonian of the trimer is diagonalized. Owing to this development it is now possible for the first time to calculate the inter- and intramolecular vibrational levels of hydrogen-bonded trimers of diatomic molecules in full dimensionality and with the degree of rigor that was so far possible only for the dimers of diatomic molecules. This methodology is used to calculate the 12D vibrational states of (HF)$_3$, encompassing the one- and two-quanta intramolecular HF-stretch excited vibrational states of the trimer and low-energy intermolecular vibrational states in the intramolecular vibrational manifolds of interest. As in I, the 2+3-body SO-3 + HF3BG PES of HF trimer\cite{QUACK98,QUACK01} is employed in these calculations. In this paper, the emphasis is on the results involving the intramolecular HF-stretch DOFs of the trimer that lie outside the scope of the rigid-monomer treatment in I. This includes the intramolecular vibrational energy levels and their frequency shifts caused by the complexation, as well as the effects of the coupling between the intra- and intermolecular vibrational modes of the trimer. Particular attention is given to the manifestations of the cooperative hydrogen bonding, due to the 3-body interactions, in the vibrational, structural, and energetic properties of HF trimer characterized by the 12D calculations.

The paper is organized as follows. Computational methodology is described in Sec. II. In Sec. III, we present and discuss the results. Section IV contains the conclusions.

\section{Computational Methodology}\label{sec_methodology}

\subsection{Coordinates, Hamiltonian, and General Approach}\label{ssection:CHGA}

\begin{figure}
\centering
\includegraphics[width=1\textwidth]{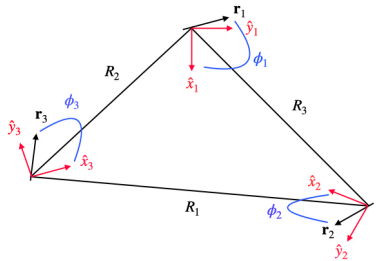}
\caption{\label{fig:coords} Schematic depiction of the coordinates used for the cyclic HF timer, assuming flexible monomers. Shown explicitly are the six in-plane coordinates: the three monomer-c.m.-to-monomer-c.m. distances $R_k$ ($k=1-3$), and the three azimuthal angles  $\phi_k$ ($k=1-3$). Also shown are the (in-plane) $\hat x_k$ and $\hat y_k$ axes of local Cartesian systems centered at the c.m. of monomer $k$ ($k=1-3$). The (out-of-plane) $\hat z_k$ ($k=1-3$) axes (not shown) are parallel to the vector ${\bf R}_1 \times {\bf R}_2$, i.e., perpendicular to the plane defined by the c.m.s of the three monomers. For each monomer $k$ the polar angle $\theta_k$ is the angle between the HF internuclear vector ${\bf r}_k$ ($k=1-3$) and the local $\hat z_k$ axis. Together, $\theta_k$ and $\phi_k$ define the orientation of ${\bf r}_k$  relative to the local Cartesian axis system attached to monomer $k$. The intramolecular HF-stretch coordinate of the $k$-th monomer is $r_k \equiv \vert {\bf r}_k \vert$. In the equilibrium geometry on the SO-3 + HF3BG PES, for $\bar r = r_e$, $R_1=R_2=R_3 = 4.76007$ bohr, $\theta_k = 90^\circ$ ($k=1-3$), and $\phi_k = 54.01979^\circ$ ($k=1-3$).  For additional details, see the text.}
\end{figure}

The full vibrational ($J=0$) Hamiltonian for HF trimer, for flexible monomers, can be written as
\begin{equation}\label{eq:1}
\hat H = \hat K_F(R)  + V'(R) + \hat K_M(R,\omega) + \hat K_{FM}(R,\omega) + \hat K_{M,rot}(\omega, r) +\hat K_{M,vib}(r) +V(R,\omega , r).
\end{equation}
Here, we use the coordinates introduced by Wang and Carrington,\cite{CARRING01B} and illustrated in Fig. \ref{fig:coords}. For the most part, they are identical to the coordinates shown in Fig. 1 of Ref. \citenum{BACIC22B}, where rigid HF monomers were assumed. Thus, $R$ represents $(R_1,R_2,R_3)$, the three coordinates giving the distances between the centers of mass of the monomers; in effect, they are the intermolecular stretching coordinates of the trimer. $\omega $ represents the $\omega _k \equiv (\theta_k, \phi_k)$, $k=1-3$, the polar and azimuthal angles, respectively, describing the orientations of the three monomer-$k$ internuclear vectors ${\bf r}_k$ (which point from the F nucleus to the H nucleus of monomer $k$) with respect to a local cartesian axis system centered at the c.m. of monomer $k$. Three new coordinates that appear in Fig. \ref{fig:coords} are the intramolecular HF-stretch coordinates  $r_k$, $k=1-3$, where $r_k \equiv \vert {\bf r}_k \vert$. The various kinetic-energy terms in Eq. (\ref{eq:1}) (all the terms but the last one in the equation) were derived by Wang and Carrington.\cite{CARRING01B} With the exception of the flexible-monomer terms they are given also in Ref. \citenum{BACIC22B} [see Eqs. (3) to (9) of that work]. The flexible-monomer terms are
\begin{equation}
 \hat K_{M,rot}(\omega, r) \equiv \sum_{k=1}^3 B_M(r_k) \hat l_k^2,
 \end{equation}
 where $B_M(r_k) \equiv 1/(2\mu_M r_k^2)$, $\mu_M$ is the reduced mass of HF monomer, and $\hat l_k^2$ is the operator associated with square of the rotational angular momentum of monomer $k$,  and 
 \begin{equation}
 \hat K_{M,vib}(r) \equiv \sum_{k=1}^3 -{1\over 2\mu_M}{\partial ^2\over \partial r_k^2}.
 \end{equation}  As mentioned above, the potential-energy function, $V$, that we use here is the SO-3 + HF3BG surface of Quack, Stohner, and Suhm.\cite{QUACK98,QUACK01}
 The volume element corresponding to $\hat H$ is of the Wilson type: $d\tau = \Pi_{k=1}^3  \sin \theta_k d\theta_k d\phi_kdR_k dr_k$.

If one defines
\begin{equation}\label{eq:Hinter}
\hat H_{\rm inter}(R,\omega;\bar r) \equiv \hat K_F(R)  + V'(R) + \hat K_M(R,\omega) + \hat K_{FM}(R,\omega) + \hat K_{M,rot}(\omega ; \bar r) + V_{\rm inter}(R,\omega;\bar r)
\end{equation}
and
\begin{equation}\label{eq:Hintra}
\hat H_{\rm intra}(r) \equiv \hat K_{M,vib}(r) + V_{\rm intra}(r;\bar R,\bar \omega )
\end{equation}
where
\begin{equation}\label{eq:Vinter}
V_{\rm inter}(R,\omega ;\bar r ) \equiv V(R,\omega ,\bar r) ,
\end{equation}
\begin{equation}\label{eq:Vintra}
V_{\rm intra}(r_1,r_2,r_3;\bar R, \bar \omega) \equiv {1\over 2}\bigg[V(\bar R, \bar \omega, r_1,r_2,r_3) + V(\bar R, \bar \omega, r_2,r_1,r_3)\bigg] ,
\end{equation}
and $\bar R$, $\bar \omega $ and $\bar r$ are collections of constants,
then
\begin{equation}\label{eq:H}
\hat H = \hat H_{\rm inter}(R,\omega;\bar r) + \hat H_{\rm intra}(r;\bar R,\bar \omega) + \Delta \hat H(R,\omega,r;\bar R, \bar \omega, \bar r).
\end{equation}
Here,
\begin{equation}
\Delta \hat H(R,\omega,r;\bar R, \bar \omega, \bar r) \equiv  \sum_{k=1}^3 \big[ B_M(r_k)-B_M(\bar r)\big]\hat l_k^2 + \Delta V(R,\omega,r;\bar R,\bar \omega ,\bar r) ,
\end{equation}
and
\begin{equation}
\Delta V(R,\omega,r;\bar R, \bar \omega , \bar r) \equiv V(R,\omega, r) - V_{\rm inter}(R,\omega;\bar r) - V_{\rm intra }(r;\bar R,\bar \omega).
\end{equation}

To solve for the eigenstates of $\hat H$ we first solve for the eigenstates of $\hat H_{\rm inter}$ and $\hat H_{\rm intra}$ in Eqs. (\ref{eq:Hinter}) and  (\ref{eq:Hintra}), respectively. The former (a 9D problem) yields eigenvectors $|I\rangle$ and corresponding eigenvalues $E_I^{\rm inter}$, while the latter (a 3D problem) yields eigenvectors $|\gamma \rangle $ and corresponding eigenvalues $E_\gamma ^ {\rm intra}$. We then construct the 12D basis states as {\bf symmetrized} products of the form 
\begin{equation}
|I,\gamma \rangle \equiv |I\rangle |\gamma \rangle ,
\end{equation}
and diagonalize the matrix of $\hat H$ in that basis.
In the $|I,\gamma \rangle $ basis the matrix elements of $\hat H$ are given by
\begin{equation}
\langle I',\gamma '|\hat H |I,\gamma\rangle = \big( E_I^{\rm inter} + E_\gamma^{\rm intra}\big)\delta _{I',I} \delta _{\gamma',\gamma} + \langle I',\gamma'|\Delta \hat H|I,\gamma\rangle.
\end{equation}
One sees that once the eigenstates of $\hat H_{\rm inter}$ and $\hat H_{\rm intra}$ are computed, the main task in constructing the $\hat H$ matrix (and, indeed, the main task in the whole 12D calculation) is to evaluate the $\langle I',\gamma '|\Delta \hat H |I,\gamma\rangle $ matrix elements. 

\subsection{Diagonalization of $\hat H_{\rm inter}$}\label{ss_Hinter}

In our first work on HF trimer\cite{BACIC22B} we present in detail the procedure employed to solve for the eigenstates of $\hat H_{\rm inter}$. We use substantially the same method here. Briefly, we first solve for the eigenstates of the 3D ``frame'' Hamiltonian, $\hat H_F(R;\bar \omega , \bar r)$, and the 6D ``bend'' Hamiltonian, $\hat H_B(\omega ; \bar R, \bar r)$. These operators are defined, respectively, in Eqs. (10) and (11) of Ref. \citenum{BACIC22B}. The eigenvectors of these operators, denoted $|\rho \rangle $ ($\rho = 1,\ldots N_F$) for the frame problem and $|\kappa \rangle $ ($\kappa = 1,\ldots , N_B$) for the bend problem are then used to construct a symmetrized product basis for the diagonalization of the 9D $\hat H_{\rm inter}$ operator, which can be expressed as
\begin{equation}
\hat H_{\rm inter}(R,\omega;\bar r) = \hat H_F(R;\bar \omega,\bar r) + \hat H_B(\omega;\bar R,\bar r) + \Delta H_{\rm inter}(R,\omega;\bar r).
\end{equation}  

\subsubsection{The frame problem}\label{sss_frame}

The 3D frame eigenvalue problem is solved by employing a basis consisting of the product of three one-dimensional potential-optimized discrete-variable representations (PODVRs) covering the $R_1$, $R_2$, and $R_3$ coordinates, respectively. The construction of the 1D PODVRs is described in Sec. II C 2 of Ref. \citenum{BACIC22B}. The 3D basis functions are of the form
\begin{equation}\label{eq:R_basis}
|n_1,n_2,n_3\rangle \equiv |R_{1,n_1}\rangle |R_{2,n_2}\rangle |R_{3,n_3}\rangle; \qquad n_k = 1,\ldots N_R
\end{equation}
where $R_{k,n_k}$ is the DVR quadrature point corresponding to the 1D PODVR function $|R_{k,n_k}\rangle $, and each $|R_{2,n}\rangle $ and $|R_{3,n}\rangle $ can be obtained, respectively, from $|R_{1,n}\rangle $ by change of variable from $R_1$ to $R_2$ and $R_1$ to $R_3$. In this work, as in Ref. \citenum{BACIC22B}, we use $N_R=12$. So, the full frame basis consists of 1,728 functions. The matrix of $\hat H_F$ in this basis is diagonalized by direct diagonalization. The $\hat H_F$ eigenvectors transform as one of the three even-parity irreducible representations (``irreps''), $A_1'$, $A_2'$, or $E'$, of the $G_{12}$ molecular symmetry group. Those pairs of degenerate eigenvectors belonging to the $E'$ irrep are then further processed (see Sec. II C 3 of Ref. \citenum{BACIC22B}) so that the resulting orthogonal pairs of states transform according to the specific representation of the $E'$ irrep that we used throughout our earlier paper.\cite{BACIC22B} One member of such a pair, the one whose eigenvalue with respect to the $(23)$ permutation operator is $-1$, belongs to what we call the $E_a'$ subirrep. The other, whose eigenvalue with respect to $(23)$ is $+1$, belongs to what we call the $E_b'$ subirrep.  

\subsubsection{The bend problem}\label{sss_bend}

The 6D bend eigenvalue problem is solved by employing a basis consisting of the product of three spherical harmonic functions covering the $\omega_1$, $\omega _2$, and $\omega _3$ local-angle coordinates:
\begin{equation}\label{eq:6Dbasis}
|l_1m_1,l_2m_2,l_3m_3\rangle \equiv Y_{l_1}^{m_1}(\theta_1,\phi_1)Y_{l_2}^{m_2}(\theta_2,\phi_2)Y_{l_3}^{m_3}(\theta_3,\phi_3),
\end{equation}
where $ l_k = 0,\ldots , l_{\rm max}$ and $m_k = -l_{\rm max},\ldots , l_{\rm max}$. In this work, as previously,\cite{BACIC22B} we use $l_{\rm max} = 13$. This gives rise to 3,766,140 even-parity basis functions and 3,763,396 odd-parity basis functions.
The matrix of $\hat H_B$ in this basis is diagonalized by a symmetry-specific version of the Chebyshev variant\cite{MANDELSTAMTAYLOR97} of filter diagonalization.\cite{NEUHAUSER95} We implement this by starting with a random initial state vector expressed in the basis of Eq. (\ref{eq:6Dbasis}) and then projecting out of that state vector that portion that corresponds to one of the eight irreps/subirreps of $G_{12}$ (i.e., $A_1'$, $A_2'$, $E_a'$, $E_b'$, $A_1''$, $A_2''$, $E_a''$, and $E_b''$.) The Chebyshev/filter-diagonalization procedure initialized with the resulting (filtered) state vector then produces eigenvectors/eigenvalues corresponding to the relevant symmetry block of $\hat H_B$. In this process repeated application of $\hat H_B$ to the state vector is required. This is accomplished by direct matrix-vector multiplication for the kinetic-energy portion of $\hat H_B$. Operation with the  potential-energy portion of $\hat H_B$ is effected by (a) transforming the state vector to a 6D Gaussian grid representation, (b) multiplying that representation of the state vector at each grid point by the value of $V_B$ at that point, and (c) transforming the result of (b) back to the original basis representation. The Gaussian grid consisted of 14 Gauss-Legendre quadrature points and 28 Fourier grid points per monomer, corresponding to the $\cos \theta_k$ and $\phi_k$ degrees of freedom, respectively, for a total grid size of 60,236,288 points.

\subsubsection{The $\hat H_{\rm inter}$ problem}\label{sss_inter}

The 9D basis in which the matrix of $\hat H_{\rm inter}$ was expressed was constructed from frame-bend product functions of the form $|\rho,\kappa\rangle \equiv |\rho\rangle |\kappa \rangle $. The $N_F$ lowest-energy frame eigenvectors of all symmetries were included in the basis and the $N_{B,\Gamma}$ lowest-energy bend eigenvectors of a each symmetry (i.e., of each irrep/subirrep) were included. We performed calculations for $N_F=150$ and 201, and for $N_{B,\Gamma}=30$ (for a total number of bend functions of each parity equal to $N_B=120$). Thus, the full 9D basis consisted of either 18,000 or 24,120 states of each parity. Symmetry-specific basis functions belonging to each of the 1D irreps and subirreps of $G_{12}$ were constructed per the procedure outlined in Sec. II E 1 of Ref. \citenum{BACIC22B} from this product basis. Matrix elements of $\hat H_{\rm inter}$ corresponding to each of the eight $G_{12}$ symmetry blocks were computed by first calculating matrix elements in the unsymmetrized $|\rho, \kappa\rangle $ basis and then transforming to the symmetrized basis. The procedures for calculating the $\hat H_{\rm inter}$ matrix elements in the $|\rho , \kappa \rangle $ basis are described in Secs. II E 2-4 of Ref. \citenum{BACIC22B}. Of these matrix elements, those corresponding to the potential-energy portion of $\Delta H_{\rm inter}$ are, by far, the most costly to compute. In this work we have reduced this load by spreading it over multiple processors (typically 30) by utilizing open-MPI. Once all the matrix elements of $\hat H_{\rm inter}$ were computed each symmetry block was diagonalized by direct diagonalization. Ultimately, the intermolecular eigenvectors, which we label $|I\rangle $, are obtained as expansions over the $|\rho , \kappa\rangle $ basis states:
\begin{equation}\label{eq:I_expansion}
|I\rangle = \sum_{\rho , \kappa }  |\rho , \kappa \rangle \langle \rho , \kappa |I\rangle 
\end{equation}

The eigenvectors and eigenvalues of $\hat H_{\rm inter} $ (as well as those of $\hat H_F$ and $\hat H_B$) depend on the value chosen for $\bar r$, the fixed intramonomer bond distance that enters into that operator as a parameter. In this work, we have performed calculations for several $\bar r$ values, each of which is different than the ones employed in Ref. \citenum{BACIC22B}. The values used herein (1.7813, 1.7843, 1.7950, and 1.8069 bohrs) were chosen with an eye toward facilitating the convergence of the 12D results. They correspond to estimates of $\langle r_k\rangle $ for the low-energy intramolecular excitations of the trimer, as obtained from the computed eigenvectors of $\hat H_{\rm intra}$ (see Sec. \ref{ss_intra_diag}). 

\subsubsection{Fixing the symmetry of the $\hat H_{\rm inter}$ eigenstates}\label{sss_inter_symmetry}

In order to make maximal use of symmetry in the 12D $\hat H$-diagonalization problem (see Sec. \ref{ss:sym1} below) one needs to work with intermolecular eigenstates that transform as the 1D irreps of $G_{12}$ and the $E_a$/$E_b$ subirreps of $G_{12}$. This requirement is automatically satisfied for the $|I\rangle $ obtained by diagonalization of the blocks of $\hat H_{\rm inter}$ that correspond to the 1D irreps. However, for those arising from the subirrep blocks (i.e., $E_a'$, $E_b'$, $E_a''$, and $E_b''$) the relation between a computed $E_a $-type intermolecular eigenvector and its corresponding computed degenerate $E_b $-type partner may not conform to our chosen $E$-type representation -- the relative phase of the two states is not fixed by their separate calculation. Hence, we need a way to guarantee that every $E_a'$, $E_b'$ (or $E_a''$, $E_b'' $) degenerate pair of intermolecular eigenstates transforms according to our choice of representation. The most straightforward way of doing this is to generate the $E_b $ eigenvector (we will call it $|b\rangle $) of any given pair directly from the $E_a$ eigenvector ($|a\rangle $) obtained from the $\hat H_{\rm inter}$ diagonalization. This is possible because for all the operations, $\hat R$, of $G_{12}$ (apart from $E^\ast$, $(23)$ and $(23)^\ast $)
\begin{equation}
|b\rangle = {\hat R|a\rangle + D_{aa}^{(E)}(\hat R)|a\rangle \over D_{ba}^{(E)}(\hat R)},
\end{equation}
where the matrix elements $D_{ij}^{(E)} (\hat R ) \equiv \langle i|\hat R |j\rangle$ ($i,j=a,b$) define the specific $E$-type representation that we desire. (These are given in Sec. I of the supplementary material of Ref. \citenum{BACIC22B}.)
Hence, one needs only to compute, for example, $(123)|a\rangle $ in order to determine $|b\rangle $ from $|a\rangle $. That calculation, in turn, requires determining how $(123)$ transforms each of the individual basis states in terms of which each $|a\rangle $ is  expressed. We treat this problem in Sec. I of the supplementary material and show in detail how each $E_b$-type intermolecular eigenvector can be computed from its $E_a$-type partner. All of the intermolecular eigenvectors of $E_b'$ and $E_b''$ symmetry that we employ in the construction of the bases used to diagonalize $\hat H$ were obtained from their $E_a'$/$E_a''$ partners by making use of these relations.

\subsection{Diagonalization of $\hat H_{\rm intra}$}\label{ss_intra_diag}

\subsubsection{Definition of the problem}\label{sss_definition}

The three-dimensional eigenvalue equation involving $\hat H_{\rm intra}$ [Eq. (\ref{eq:Hintra})] is specified once the constants $\bar R$ and $\bar \omega $ that define $V_{\rm intra}$ [Eq. (\ref{eq:Vintra})] are specified. We choose $\bar R_k = 5.0$ bohrs, $\bar \omega _k \equiv (\bar \theta_k, \bar \phi_k) = (90^\circ, 60^\circ)$, $k=1-3$. These values are close to the vibrationally averaged values of $R_k$ and $\omega _k$ associated with the ground state of $\hat H_{\rm inter}$ (see Ref. \citenum{BACIC22B}). Hence $V_{\rm intra}$ represents a ``trimer-adapted'' potential for the collective intramonomer motions in the trimer, rather than a sum of three isolated-monomer potentials. By taking this approach we anticipate that a substantial part of the effect of intermonomer interactions on the intramonomer vibrations in the trimer can be captured in the 3D $|\gamma\rangle $ eigenvectors. It should also be noted that the average that appears on the rhs of Eq. (\ref{eq:Vintra}), the definition of $V_{\rm intra}$, renders that function invariant with respect to the operations of $G_{12}$, the molecular symmetry group characterizing HF trimer. Hence, $\hat H_{\rm intra}$ is also invariant with respect to those operations, and each of its eigenvectors transforms according to one of the irreps of $G_{12}$. 

\subsubsection{Primitive intramolecular basis and matrix elements}\label{sss_primitive_basis}

To solve
\begin{equation}\label{eq:HintraSE}
\hat H_{\rm intra} |\gamma \rangle = \bigg[ \sum_{k=1}^3 \bigg( -{1\over 2\mu_M} {\partial ^2\over \partial r_k^2} \bigg) + V_{\rm intra}(r)\bigg] |\gamma \rangle =  E_\gamma^{\rm intra}|\gamma\rangle
\end{equation}
we work in a 3D potential-optimized discrete-variable-representation (PODVR) basis consisting of product functions of the form
\begin{equation}\label{eq:rmDVR}
|j\rangle \equiv |j_1,j_2,j_3\rangle \equiv |{\bf r}_j\rangle \equiv |r_{1,j_1}\rangle |r_{2,j_2}\rangle |r_{3,j_3}\rangle ; \qquad j_k = 1,\ldots , N_r 
\end{equation}
where the meta index $j$ represents $(j_1,j_2,j_3)$ and the $|r_{k,j_k}\rangle $ are 1D PODVR functions. The latter are derived from the $N_r$ lowest-energy eigenvectors of the 1D equation
\begin{equation}\label{eq:rmPODVReq}
\bigg[ -{1\over 2\mu_M}{\partial ^2\over \partial r_k^2}  + V_{1D}(r_k) \bigg] |f\rangle = E |f \rangle ,
\end{equation}
where $V_{1D}(r_k)$ is the potential obtained from $V_{\rm intra}$ by fixing the two $r_i$ ($i\ne k$) coordinates to identical constants corresponding to a value close to the monomer ground-state bond-distance expectation value. These energy eigenvectors were computed by direct diagonalization of Eq. (\ref{eq:rmPODVReq}) in a basis of 200 sinc-DVR functions corresponding to $r_k$ quadrature points ranging from 1 to 3.2 bohrs.  The PODVR functions were then obtained by diagonalizing the matrix of $r_k$ in the energy-eigenvector basis. Each of the 1D PODVR functions is denoted by the $r_k$ eigenvalue (i.e., the PODVR quadrature point $r_{k,j_k}$) to which it corresponds. Due to the permutation symmetry characterizing the HF trimer the determination of the PODVR functions for $k=1$, say, yields those for the other two simply by replacing the coordinate $r_1$ in those functions with $r_2$ and $r_3$, respectively. 

The matrices in the PODVR basis of each of the three terms in the kinetic-energy portion of $\hat H_{\rm intra}$ were computed by first evaluating the matrix elements of $(\partial ^2/\partial r_k^2)$ in the 1D particle-in-a-box basis from which the sinc-DVR basis used to solve Eq. (\ref{eq:rmPODVReq}) was derived. These matrix elements were then transformed to the sinc-DVR basis and, finally, to the PODVR basis. The matrix elements of $V_{\rm intra}$ in the PODVR basis were evaluated by quadrature. The $\hat H_{\rm intra}$ matrix in the PODVR basis was diagonalized directly. The $\hat H_{\rm intra}$ eigenvectors are obtained as expansions over the primitive intramolecular basis
\begin{equation}\label{eq:intra_evecs}
|\gamma \rangle = \sum _j |j\rangle \langle j|\gamma\rangle = \sum_{j_1,j_2,j_3} | j_1,j_2,j_3\rangle \langle j_1,j_2,j_3|\gamma\rangle
\end{equation}

To obtain the $|\gamma\rangle $ employed in constructing the 12D bases we primarily used $N_r=8$ and $N_r=10$. Hence, the 3D primitive basis consisted of 512 PODVR functions for $N_r=8$ and 1000 PODVR functions for $N_r=10$. This basis is the smallest that produces reasonable convergence of the intramolecular eigenenergies corresponding to the 3D ground state (converged to within 0.01 cm$^{-1}$) and the $v=1$ (converged to a few 0.1 cm$^{-1}$) and $v=2$ (converged to a few cm$^{-1}$) excited states. Since the 12D computational cost scales as $(N_r)^3$ (see below), it is advantageous to keep $N_r$ as small as is reasonable. In order to assess the convergence of the 3D eigenstates we also performed calculations for which $N_r=12$ and 16. 

\subsubsection{Symmetry considerations}\label{sss_symmetry_consider}

An eigenstate of $\hat H_{\rm intra}$ transforms as one of the even-parity irreps of $G_{12}$: $A_1'$, $A_2'$, or $E'$. As with the frame, bend, and $\hat H_{\rm inter}$ doubly-degenerate eigenstates, we find it convenient to work with doubly-degenerate $\hat H_{\rm intra}$ eigenstates that transform like the specific $E$-type representation detailed in Sec. I of the supplementary material of Ref. \citenum{BACIC22B}. Hence, doubly-degenerate $E'$ eigenvector pairs produced by the diagonalization of $\hat H_{\rm intra}$ were subjected to further processing so as to fix their transformation properties with respect to the operations of $G_{12}$.  For any given degenerate pair of states obtained by solving Eq. (\ref{eq:rmPODVReq}) we accomplish this by diagonalizing the matrix of the $(23)$ permutation operator in the basis consisting of that pair. We then choose (a) the resulting $(23)$ eigenstate having eigenvalue $-1$ as one of the states with the desired transformation properties (the $E_a^\prime$ member of the pair) and (b) the other state, with $(23)$ eigenvalue of $+1$, to be the other state (the $E_b'$ member of the pair) after adjusting its phase so that it has the correct transformation properties with respect to the $(123)$ permutation operation. (This procedure is analogous to that described in greater detail for the frame eigenstates in Sec. II C 3  of Ref. \citenum{BACIC22B}.) The result is a pair of orthogonal, degenerate $E'$ states that transform according to our desired representation.

\subsection{Diagonalization of $\hat H$}\label{ss_H_diag}

\subsubsection{The symmetrized 12D basis}\label{sss:12Dbasis}

As mentioned in Sec. \ref{ssection:CHGA}, the basis we use to diagonalize $\hat H$ consists of symmetrized functions composed of products of the form $|I,\gamma\rangle \equiv |I\rangle |\gamma\rangle $. In the case where either $|I\rangle$ and/or $|\gamma \rangle $ belongs to a 1D irrep of $G_{12}$, the symmetrized function constructed from $|I,\gamma\rangle $ is $|I,\gamma\rangle $ itself. However, if both $|I\rangle $ and $|\gamma \rangle $ are members of a doubly-degenerate pair, then the symmetrized basis functions that are constructed from the four relevant $|I,\gamma\rangle $ belong, respectively, to the irreps $A_1^s$, $A_2^s$, and subirreps $E_a^s$ and $E_b^s$ ($s = \prime$ or $\prime\prime$ depending on whether $|I\rangle $ is of even or odd parity, respectively). These symmetrized basis functions are of the form
\begin{equation}
|I,\gamma\rangle _{A_1^s} = \sqrt{1\over 2} \bigg[ |I_a,\gamma_a\rangle + |I_b,\gamma _b\rangle \bigg],
\end{equation}
\begin{equation}
|I,\gamma\rangle _{A_2^s} = \sqrt{1\over 2} \bigg[ |I_a,\gamma_b\rangle - |I_b,\gamma _a\rangle \bigg],
\end{equation}
\begin{equation}
|I,\gamma\rangle _{E_a^s} = \sqrt{1\over 2} \bigg[ |I_a,\gamma_b\rangle + |I_b,\gamma _a\rangle \bigg],
\end{equation}
and
\begin{equation}
|I,\gamma\rangle _{E_b^s} = \sqrt{1\over 2} \bigg[ |I_a,\gamma_a\rangle - |I_b,\gamma _b\rangle \bigg].
\end{equation}

From all this one sees that the $\hat H_{\rm inter} + \hat H_{\rm intra}$ portion of $\hat H$ [see Eq. (\ref{eq:H})] is diagonal in the symmetrized basis with matrix elements that are simply sums of intermolecular and intramolecular eigenvalues. The matrix elements of $\Delta \hat H$ in the symmetrized 12D basis are easily obtained from the matrix elements of $\Delta \hat H$ in the $|I,\gamma\rangle $ basis. We concentrate on the calculation of these quantities in what follows.

\subsubsection{$\Delta \hat H$ matrix elements: General}\label{sss_deltaH_general}

In order to calculate matrix elements of the form $\langle I',\gamma '|\Delta \hat H|I,\gamma\rangle $ we first express the $|I\rangle $ in terms of the frame ($|\rho \rangle $), bend ($|\kappa \rangle $) basis states $|\rho, \kappa\rangle $ [Eq. (\ref{eq:I_expansion})]
and the $|\gamma\rangle $ in terms of the 3D DVR states [Eq. (\ref{eq:intra_evecs})].
Then, the kinetic-energy term in the $\Delta \hat H$ matrix element is given by
%\end{document}
\begin{eqnarray}\label{eq:DHKEO}
\langle I',\gamma'|\bigg[\sum_{k=1}^3 \big[ B_M(r_k) - B_M(\bar r)\big]\hat l_k^2 \bigg] |I,\gamma \rangle &=& \sum_{\kappa',\kappa}\sum_{\rho}\sum_{j} \bigg[\sum_{k=1}^3 \langle \kappa '| \hat l_k^2 |\kappa\rangle \big[ B_M(r_{k,j_k})-B_M(\bar r)\big] \bigg] \nonumber \\ && \times \langle I'|\rho,\kappa'\rangle \langle \rho, \kappa  | I\rangle \langle \gamma ' |j\rangle \langle j|\gamma \rangle. \nonumber \\
\end{eqnarray}
The potential-energy term is given by
\begin{eqnarray}\label{eq:DHPEO}
\langle I',\gamma '| \Delta V(R,\omega,r)|I,\gamma\rangle &=& \sum_{\kappa',\kappa}\sum_{\rho',\rho}\sum_j \langle \rho', \kappa',j|\Delta V(R,\omega , r) |\rho, \kappa , j\rangle \nonumber \\
&&\times  \langle I'| \rho ',\kappa '\rangle \langle \rho, \kappa  | I \rangle \langle \gamma '|j\rangle \langle j|\gamma\rangle       \nonumber \\  &=& \sum_{\kappa',\kappa}\sum_{\rho',\rho}\sum_j \langle \rho ',\kappa' | \Delta V(R,\omega , {\bf r}_j) |\rho, \kappa \rangle \nonumber \\
&&\times  \langle I'| \rho ',\kappa '\rangle \langle \rho , \kappa  | I \rangle \langle \gamma '|j\rangle \langle j|\gamma\rangle.       \nonumber \\
\end{eqnarray}

\subsubsection{Computing the matrix elements of the kinetic-energy portion of $\Delta \hat H$}\label{sss_deltaH_kinetic}

The matrix elements on the rhs of Eq. (\ref{eq:DHKEO}) can be evaluated by expressing the $|\kappa\rangle $ in terms of the primitive basis states $|l_1m_1,l_2m_2,l_3m_3\rangle $ [Eq.(\ref{eq:6Dbasis})]. One has
\begin{equation}\label{eq:lksq}
\langle \kappa'|\hat l_k^2|\kappa \rangle = \sum_{l,m} l_k(l_k+1) \langle \kappa'|lm\rangle \langle lm|\kappa\rangle 
\end{equation}
with $|lm\rangle \equiv |l_1m_1,l_2m_2,l_3m_3\rangle $. One can also define the quantities
\begin{equation}\label{eq:DelB}
\langle \gamma '|\Delta B(k) |\gamma\rangle \equiv \sum_{j_1,j_2,j_3} \langle \gamma' |j_1,j_2,j_3\rangle \langle j_1,j_2,j_3|\gamma \rangle \big[ B_M(r_{k,j_k})-B_M(\bar r)\big]
\end{equation}
so that Eq. (\ref{eq:DHKEO}) becomes
\begin{eqnarray}\label{eq:DHKEO2}
\langle I',\gamma'|\bigg[\sum_{k=1}^3 \big[ B_M(r_k) - B_M(\bar r)\big]\hat l_k^2 \bigg] |I,\gamma \rangle &=& \sum_{\kappa',\kappa}\sum_{\rho}\bigg[\sum_{k=1}^3 \langle \kappa '| \hat l_k^2 |\kappa\rangle \langle \gamma' |\Delta B(k)|\gamma \rangle  \bigg] \nonumber \\ &&\qquad\qquad\qquad\qquad \times \langle I'|\rho ,\kappa'\rangle \langle \rho ,\kappa |I\rangle. \nonumber \\
\end{eqnarray}
All of the required $\langle \kappa'|\hat l_k^2|\kappa\rangle $ [Eq. (\ref{eq:lksq})] and the $\langle \gamma '|\Delta B(k)|\gamma \rangle $ [Eq. (\ref{eq:DelB})] can be computed separately prior to the evaluation of Eq. (\ref{eq:DHKEO2}), and such evaluation for all $I'$, $I$, $\gamma '$, and $\gamma $ is ultimately not especially expensive.

\subsubsection{Computing the matrix elements of $\Delta V$: General procedure}\label{ss:GenDV}\label{sss_deltaV_general}

The $\Delta V$ matrix elements that appear on the rhs of Eq. (\ref{eq:DHPEO}) are identical to those that enter into the 9D (i.e., intermolecular) trimer problem. They are readily evaluated by expressing the $|\rho \rangle $ in terms of the primitive 3D DVR basis states $|n\rangle \equiv |n_1,n_2,n_3\rangle \equiv |R_{1,n_1}\rangle |R_{2,n_2}\rangle |R_{3,n_3}\rangle $ [Eq. (\ref{eq:R_basis})], which gives
\begin{equation}\label{eq:V9Dme}
 \langle \rho ', \kappa'|\Delta V(R,\omega , {\bf r}_j) |\rho ,\kappa \rangle = \sum_n \langle \kappa '|\Delta V({\bf R}_n,\omega , {\bf r}_j)|\kappa\rangle \langle \rho '|n\rangle \langle n|\rho\rangle.
 \end{equation}
 One sees that it is necessary to evaluate 6D integrals over an $\omega $ grid at each of the $R,r$ DVR points. That is, one needs to compute the quantities
 \begin{equation}\label{eq:DelVkappa}
\Delta  V^{\kappa',\kappa}_{n,j}\equiv \langle \kappa '|\Delta V({\bf R}_n,\omega , {\bf r}_j)|\kappa \rangle = \sum_{i=1}^{N_\omega} \langle \kappa'|\omega_i\rangle \Delta V({\bf R}_n, \omega _i, {\bf r}_j)\langle \omega _i |\kappa \rangle 
 \end{equation}
 for all $j$, $n$, $\kappa $, and $\kappa '$. (This is a variation of the ``F-matrix'' approach to the calculation of PES matrix elements introduced by Carrington, et al.\cite{CARRING94,CARRING02})
 With Eqs. (\ref{eq:V9Dme}) and (\ref{eq:DelVkappa}), Eq. (\ref{eq:DHPEO}) becomes
 \begin{eqnarray}\label{eq:DHPEO2}
\langle I',\gamma '| \Delta V(R,\omega,r)|I,\gamma\rangle &=&\sum_j \langle \gamma '|j\rangle \langle j|\gamma\rangle  \nonumber \\  &&\qquad \times  \sum_{\kappa',\kappa}  \sum_n  \Delta V_{n,j}^{\kappa',\kappa}  \sum_{\rho',\rho} \langle I'| \rho ', \kappa '\rangle \langle \rho '|n\rangle  \langle n|\rho\rangle \langle \rho ,\kappa  | I \rangle.
\end{eqnarray}

We evaluate Eq. (\ref{eq:DHPEO2}) as follows. First, in a loop over $n=1$ to $n_{\rm max}$ for initial value $j=1$, we (1) compute the PES values $\Delta V({\bf R}_n,\omega _i , {\bf r}_j)$  for all $i$, (2) compute $\Delta V_{n,j}^{\kappa ',\kappa}$ for all $\kappa', \kappa$, and (3) accumulate the quantities 
\begin{equation}
F_{n,j}^{I',I} = F_{n-1,j}^{I',I } + \sum_{\kappa',\kappa} \Delta V_{n,j}^{\kappa ',\kappa } \langle I'|n,\kappa '\rangle \langle n, \kappa|I \rangle ;\qquad  F_{0,j}^{I',I }=0 
\end{equation}
for all $I'$, $I$  as the $n$ loop progresses. Here, the
\begin{equation}
\langle n,\kappa |I\rangle \equiv \sum_\rho \langle n|\rho \rangle \langle \rho, \kappa | I\rangle 
\end{equation}
are quantities that are only calculated and stored as needed for the current value of $n$ in the $n$ loop.
With the $n$ loop finished ($n=n_{\rm max}$) one has computed the quantities
\begin{equation}
G^{I',I}_j \equiv F^{I',I}_{n_{\rm max},j} =  \sum_n  \sum_{\kappa',
\kappa} \Delta V_{n,j}^{\kappa',\kappa} \langle I '|n,\kappa '\rangle \langle n,\kappa |I\rangle =\langle I',j|\Delta V(R,\omega,r)|I,j\rangle
\end{equation}
for all $I'$, $I$ for the specific value of $j$. This completed, we advance to the next $j$ and repeat the whole procedure. This process continues until the $G^{I',I}_j$ for all $I'$, $I$ are computed for all $j$.
Finally, we compute the desired matrix elements
\begin{equation}
 \langle I',\gamma'|\Delta V(R,\omega,r)|I,\gamma\rangle = \sum_j G_j^{I',I} \langle \gamma'|j\rangle \langle j|\gamma \rangle.
\end{equation}

\subsubsection{Computing the matrix elements of $\Delta V$: Exploitation of symmetry}\label{ss:sym1}

The intermolecular eigenstate $|I\rangle $ has either positive or negative parity, while all the intramolecular eigenstates $|\gamma\rangle $ have positive parity. As such, and since $\Delta V(R,\omega,r)$ is invariant with respect to inversion, $\langle I',\gamma'|\Delta V(R,\omega,r)|I,\gamma\rangle = 0$ unless $|I\rangle $ and $|I'\rangle $ have the same parity. Thus, one only needs to work through the whole $ \langle I',\gamma'|\Delta V(R,\omega,r)|I,\gamma\rangle$ evaluation process for intermolecular states that have the same parity.  Further, since the frame eigenstates $|\rho\rangle $ all have positive parity, the only bend eigenstates, $|\kappa \rangle $, that contribute to any given $|I\rangle $ must have the same parity as that $|I\rangle$. In consequence, there is no need to evaluate $\Delta V_{n,j}^{\kappa', \kappa}$ quantities for which $|\kappa '\rangle $ and $|\kappa \rangle $ have different parity (indeed such quantities equal zero). The upshot of all this is that the procedure of Sec. \ref{ss:GenDV} can be made parity-specific without loss of any accuracy. This allows for a reduction by about a factor of two in the cost of the entire 12D calculation. 

In the procedure of Sec. \ref{ss:GenDV} one can also use the permutation-symmetry operations of $G_{12}$ to reduce the number of iterations in the $j$ loop by focusing on the quantities $G^{I',I}_j$ produced in each iteration of that loop. One has
\begin{eqnarray}
G^{I',I}_j = \langle I',j|\Delta V(R,\omega , r)|I,j\rangle &=&  \langle I',j|\hat R^{-1}[\hat R\Delta V(R,\omega , r)\hat R^{-1}]\hat R|I,j\rangle \nonumber \\ &=&\langle I',j|\hat R^{-1}[\Delta V(R,\omega , r)]\hat R|I,j\rangle, \nonumber \\
\end{eqnarray}
where $\hat R$ is an operation of $G_{12}$, and we have used the invariance of the $\Delta V$ function to such operations. Now, note that
\begin{equation}
\hat R|I,j\rangle = \hat R|I\rangle \times \hat R|j_1,j_2,j_3\rangle.
\end{equation}
For the permutation operations of $G_{12}$, the functions $\hat R |j_1,j_2,j_3\rangle $ are easily obtained. The $\hat R|I\rangle $, in turn, are given by
\begin{equation}
\hat R|I\rangle  = \chi^{(\Gamma_I)}(\hat R)|I\rangle ,
\end{equation}
where $\chi^{(\Gamma_I)}(\hat R)$ is the character corresponding to $\hat R$ for 1D irrep $\Gamma_I$,  if $|I\rangle$ is singly-degenerate,
and by
\begin{equation}
\hat R|I_k\rangle = \sum_{k'= a,b} D^{(\Gamma_I)}_{k',k}(\hat R) |I_{k'}\rangle ;\qquad \qquad k=a,b
\end{equation}
if it is one of a doubly-degenerate pair. 

The upshot is that one can express the $G^{I',I}_j$ for one value of $j$ in terms of those evaluated for another $j$ value. There are three cases. When $|I\rangle $ and  $|I'\rangle $ are both singly-degenerate then
\begin{equation}
G^{I',I}_j \equiv G^{I',I}_{(j_1,j_2,j_3)}  = \chi^{(\Gamma_{I'})} (\hat R) \chi^{(\Gamma_I)}(\hat R) G^{I',I}_{\hat R (j_1,j_2,j_3)}
\end{equation}
or, equivalently,
\begin{equation}\label{eq:Gtransform1}
G^{I',I}_{\hat R(j)} \equiv G^{I',I}_{\hat R(j_1,j_2,j_3)}  = \chi^{(\Gamma_{I'})} (\hat R^{-1}) \chi^{(\Gamma_I)}(\hat R^{-1}) G^{I',I}_{(j_1,j_2,j_3)}.
\end{equation}
When $|I'\rangle $ is doubly-degenerate and $|I\rangle $ is nondegenerate, then one has
\begin{equation}\label{eq:Gtransform2}
G^{I_k',I}_{\hat R(j)} \equiv G^{I'_k,I}_{\hat R(j_1,j_2,j_3)}  = \sum_{k'=a,b}D_{k',k}^{(\Gamma_{I'})} (\hat R^{-1}) \chi^{(\Gamma_I)}(\hat R^{-1}) G^{I'_{k'},I}_ {(j_1,j_2,j_3)}; \qquad\qquad k=a,b
\end{equation}
Hermiticity can then be used to obtain
\begin{equation}\label{eq:Gtransform1a}
G^{I,I_k'}_{\hat R(j)} = \big[G^{I_k',I}_{\hat R(j)}  \big]^\ast.
\end{equation}
Finally, when both $|I\rangle $ and $|I'\rangle $ are doubly-degenerate
\begin{equation}\label{eq:Gtransform3}
G^{I_k',I_m'}_{\hat R(j)}  \equiv G^{I'_k,I_m'}_{\hat R(j_1,j_2,j_3)}  = \sum_{k'=a,b}~\sum_{m'=a,b}D_{k',k}^{(\Gamma_{I'})} (\hat R^{-1}) D_{m',m}^{(\Gamma_I)}(\hat R^{-1}) G^{I'_{k'},I_{m'}}_{ (j_1,j_2,j_3)}; \qquad k=a,b,\quad m=a,b
\end{equation}
It turns out that there are $\bar N_j\equiv N_r(N_r+1)(N_r+2)/6$ sets of $j=(j_1,j_2,j_3)$ triplets, within which sets the triplets are transformed into one another by the permutation operations of $G_{12}$. Hence, given Eqs. 
(\ref{eq:Gtransform1}) to (\ref{eq:Gtransform3}), the $j$ loop need only encompass $j$ values consisting of just one representative of each such set. The $G^{I',I}_j$ quantities not computed directly in the loop can then be obtained from those that are from Eqs. 
(\ref{eq:Gtransform1}) to (\ref{eq:Gtransform3}). Thus, symmetry can be used to reduce the iterations of the $j$ loop by a factor of $N_j/\bar N_j=6N_r^3/[N_r(N_r+1)(N_r+2)]$. For the $N_r=8$ and $N_r=10$ values employed here this reduction is by a factor of about 4.27 and 4.55, respectively.

A further reduction in computational steps can be effected by choosing to deal with the following representatives of the $\bar N_j$ symmetry-connected sets of $j$ triplets: (1) type-1 triplets of the form $j_1=j_2=j_3$, of which there are $\bar N_j^{(1)}=N_r$, are the sole members of the symmetry-connected sets for which all three of the $j_k$ values are equal, (2) type-2 triplets of the form $j_1\ne j_2$, $j_2=j_3$, of which there are $\bar N_j^{(2)}=N_r(N_r-1)$, cover all of the symmetry-connected sets for which precisely two of the $j$ values are the same, and (3) type-3 triplets of the form $j_1<j_2<j_3$, of which there are $\bar N_j^{(3)}=N_r(N_r-1)(N_r-2)/6$, cover all of the symmetry-connected sets of $j$ for which $j_1$, $j_2$ and $j_3$ are all different. Now, when the value of $j$ in the $j=1$-to-$\bar N_j$ loop is type-3, there is nothing further to be gained by making use of symmetry than what has already been described. However, when $j$ is type-1 or type-2, symmetry can be used to reduce the number of direct calculations of $\Delta V_{n,j}^{\kappa ',\kappa}$ values [Eq. (\ref{eq:DelVkappa})] in the $n$ loop. We present in Sec. II of the supplementary material the details of how this can be done and show therein that the total number of iterations required in the nested $j$ and $n$ loops when permutation symmetry is exploited is $N_rN_R(N_rN_R+1)(N_rN_R+2)/6$, which compares to the $(N_rN_R)^3$ iterations required when such symmetry is neglected. In short, the speed-up obtainable by exploiting permutation symmetry is almost a factor of six.

\subsubsection{Computing the matrix elements of $\Delta V$: Limiting the size of the $\omega_i$ grid}\label{sss_deltaV_grid}

In the method described above the calculation of all the required $\Delta V^{\kappa',\kappa}_{n,j}$ represents, by far, the largest cost in the entire 12D calculation. The number of such evaluations is on the order of $[N_B (N_B +1)/2]\times [N_rN_R(N_rN_R+1)(N_rN_R+2)/6] $. With the typical values $N_B = 120$, $N_r = 8$ and $N_R=12$ this amounts to about $1.1\times 10^9$ evaluations for a given parity. And, each such evaluation requires a sum over $N_\omega$ grid points. Such a grid must nominally consist of $N_\omega= (266)^3\simeq 1.9\times 10^7$ Lebedev points to accommodate the $l_{\rm max}=13$ primitive angle basis that we require to compute the converged bend states $|\kappa\rangle$. In short, even when permutation symmetry is fully exploited one is nominally faced with sums over a total of about $2\times 10^{16}$ points in order to compute all of the $\Delta V^{\kappa',\kappa}_{n,j}$ required. A reduction in this number of operations by a significant factor reduces the cost of the overall 12D calculation by about the same factor. Hence, one is interested in trying to effect such a reduction. 

Elsewhere\cite{BACIC22B} we have shown that a reduction in the size of the 6D angle grid ($N_\omega$) by about a factor of two can, by making use of parity, be effected without loss of any accuracy. Here, we introduce an approximation leading to a further significant reduction in the $\omega_i$ grid size. The approximation relies on the assumption that for many $i$ the $\langle \kappa'|\omega_i\rangle \langle \omega_i|\kappa\rangle $ quantities that enter into Eq. (\ref{eq:DelVkappa}) are vanishingly small for all $\kappa, \kappa '$. If this is true, then these grid points can be neglected in every evaluation of Eq. (\ref{eq:DelVkappa}) and $N_\omega$ can be effectively reduced. The physical basis for this approximation is the evidence that the wavefunctions of the low-energy bend excitations in HF trimer are substantially localized in angle space,\cite{BACIC22B} so that nowhere near the entire 6D space is required to produce accurate grid representations of them.

In order to implement this approximation one needs an efficient way to search for those $\omega _i$ for which $|\langle \omega _i|\kappa\rangle |$ is very small for all $\kappa$. We do this as follows. We construct the aggregate probability density 
\begin{equation}
\rho_{\rm tot}(\omega_i) \equiv \sum_{\kappa=1}^{N_B} |\langle \omega_i|\kappa\rangle |^2
\end{equation}
at every grid point. We then sort the grid points into those for which
\begin{equation}
\rho_{\rm tot}(\omega_i) \ge {A\over N_B N_\omega},
\end{equation}
where $A$ is an adjustable, dimensionless constant of the order 0.1. We retain these grid points in computing Eq. (\ref{eq:DelVkappa}) and discard all of the others. We can get a sense of the accuracy of this procedure by computing the norm of each of the bend eigenstates over this trimmed grid
\begin{equation}
\langle \kappa |\kappa\rangle_{\rm trim}\equiv \sum_{i_{\rm trim}=1}^{N_{\omega_{\rm trim}}} |\langle \omega_{i_{\rm trim}}|\kappa\rangle |^2
\end{equation}
and comparing these values to unity. We find that for $A=0.1$ all of these trimmed norms for the set of $|\kappa\rangle $ that we retain in the $\hat H_{\rm inter}$ bases differ by less than 3 parts in $10^4$ from 1. Further, at this level of approximation the trimmed grid size is only about 7\% of the size of the exact parity-reduced grid. By reducing $A$ by a factor of five to $0.02$ all of the trimmed norms for the set of $|\kappa \rangle $ that we use differ by less than 7 parts in $10^5$ from unity, and the trimmed grid size is about 10\% of the full grid. 

Of course, the rigorous test of this trimming approximation is an assessment of how the magnitude of $A$ affects the solutions to the 12D problem. We find a negligible increase (of order 0.01 cm$^{-1}$) in computed 12D eigenenergies in going from $A=0.1$ to $A=0.02$. As such, we are confident that the $A=0.1$ results are an accurate approximation to those associated with the untrimmed grid.

\subsection{12D Expectation values of geometrical quantities}\label{ss_12D_expectationvalues}

In examining the nature of the eigenstates of $\hat H$ it is of interest to compute the expectation and root-mean-squared values of various geometrical quantities, e.g., intra- and intermolecular bond lengths, distances, and angles. Such computations are nontrivial, but straightforward, given that the 12D eigenstates are obtained as expansions over products of lower-dimension functions. We outline the calculation of such quantities in detail in Sec. III of the supplementary material.

\section{Results and discussion}\label{s_results}

\subsection{Convergence tests}\label{ss_convergence}

Parameters of the basis sets used in the 12D calculations utilizing the methodology of Sec. \ref{sec_methodology} were carefully tested for convergence. Eight different 12D basis sets representative of the convergence testing performed are listed Table \ref{tab_12D_bases}. As mentioned in Sec. \ref{sss_inter}, the eigenvectors and eigenvalues of $\hat H_{\rm inter}$ depend on the value chosen for $\bar r$, the rigid-HF monomer bond length that appears in that operator as a parameter. While in the limit of a very large 9D intermolecular basis the final 12D results would be insensitive to the $\bar r$ value, its judicious choice can facilitate their convergence. The $\bar r$ values  in Table \ref{tab_12D_bases} are 1.7950 bohrs for bases I/IA - IV, 1.7843 bohrs for bases V and VI, and 1.7813 bohrs for basis VII. The rationale for choosing these values and the effect they have on the 12D results are presented below. In the first two groups of bases, the numbering of the bases increases with decreasing $N_{\rm inter}$, from 600 to 396 for bases I/IA - IV, and from 600 to 240 for bases V-VI. Different bases are further distinguished by $N_F$, the total number of 3D frame states (Sec. \ref{sss_frame}) included in the 9D intermolecular basis. $N_r$, the number of 1D PODVR basis functions per HF-stretch coordinates $r_k$ $(k=1-3)$ used in diagonalizing the $\hat H_{\rm intra}$ matrix, is set to 8 in all bases except basis IA, where $N_r=10$. The significance of this is discussed below. For all bases in Table \ref{tab_12D_bases} $N_B$, the total number of 6D bend states (Sec. \ref{sss_bend}) of a given parity used to build the 9D basis, is set to 120, or 30 bend states per irrep.  In addition, for all bases, $N_{intra}$ is equal to 56, which encompasses all 3D eigenstates of $\hat H_{\rm intra}$ with up to and including five quanta of HF stretch excitation. Test 12D calculations were performed with $N_{intra}=35$ (which covers all intramolecular states up to and including four HF-stretch quanta)  and negligible difference was found from the results for $N_{intra}=56$. 

Inspection of Table \ref{tab_12D_intra} reveals that increasing  $\bar r$ from 1.7843 bohrs in basis V to 1.7950 bohrs in basis I, while keeping all other basis-set parameters the same, has a small but visible effect on the intramolecular $v=1,2$ HF stretching states in HF trimer. The energies of the $v=1$ HF-stretch states, $\nu^{\rm HF}_{\rm sym}$ and $\nu^{\rm HF}_{\rm asym}$, from the 12D calculations decrease only by about 0.4 cm$^{-1}$. But, for the four states with two quanta in HF-stretch modes of the trimer, including $2\nu^{\rm HF}_{\rm sym}$ and $\nu^{\rm HF}_{\rm sym}+ \nu^{\rm HF}_{\rm asym}$, their energies computed using basis I are 1.2-1.9 cm$^{-1}$ lower than those obtained with basis V. This should not be surprising. The $\bar r = 1.7950$ bohrs value in basis I was chosen to be between the expectation values of the HF monomer bond length for the ground state and the intramolecular symmetric-stretch fundamental of the HF trimer, 1.7843 and 1.8069 bohrs, respectively, from the reduced-dimension quantum 3D calculations of the intramolecular vibrational states of (HF)$_3$ (Sec. \ref{ss_intra_diag}). In contrast, the $\bar r$ value of  1.7843 bohrs used in basis V is equal to that of the quantum 3D expectation value for ground state of the trimer. Therefore, the larger value of  $\bar r$ in basis I reflects better the increase of the vibrationally averaged HF bond length upon intramolecular vibrational excitation than the smaller $\bar r$  value used in basis V, making it more appropriate for the HF-stretch excited states of the trimer. It should be added that quantum 12D calculations were performed also for $\bar r = 1.8069$ bohrs, equal to the expectation value of the HF bond length for the intramolecular symmetric-stretch fundamental of the HF trimer, with all other basis-set parameters as in basis I. The energies of the $v=1$ HF-stretch states of the trimer they yielded are about 0.4 cm$^{-1}$ lower than those obtained for $\bar r = 1.7950$ (basis I), while those of the $v=2$ trimer states decrease by about 1 cm$^{-1}$. But, the energies of some intermolecular bending states for $\bar r = 1.8069$ bohrs are 1-2 cm$^{-1}$ higher than those computed with $\bar r = 1.7950$. Given this mixed performance, $\bar r = 1.8069$ bohrs was given no further consideration.

Table \ref{tab_12D_intra} also shows that the 12D energies of the $v=1$ HF-stretch states of HF trimer, $\nu^{\rm HF}_{\rm sym}$ and $\nu^{\rm HF}_{\rm asym}$, hardly change for bases I-IV, which all have $\bar r = 1.7950$ but differ in the value of $N_{\rm inter}$, from 600 (basis I)  to 396 (basis IV), implying that these states are well-converged with respect to this important parameter. For the same bases I-IV,  increasing $N_{\rm inter}$ from 396 to 600 lowers the energy of the $v=2$ trimer state $2\nu^{\rm HF}_{\rm sym}$ by no more than 0.70 cm$^{-1}$.

Table \ref{tab_12D_converge} displays the basis-set dependence of the selected low-energy intermolecular vibrational states of HF trimer from 12D quantum calculations, for the monomers in their ground intramolecular vibrational state. Comparison of the results obtained for bases I and V, respectively, shows that they change very little in going from $\bar r = 1.7950$ bohrs to $\bar r = 1.7843$ bohrs. Furthermore, the differences between the energy levels calculated for bases I-IV are generally very small, although $N_{\rm inter}$ changes from 396 to 600. In particular, the results obtained using basis I ($N_{\rm inter}=600$) and basis II ($N_{\rm inter}=540$) differ by about 0.1 cm$^{-1}$, demonstrating good convergence with respect to $N_{\rm inter}$.

The 12D results in Tables \ref{tab_12D_intra} and \ref{tab_12D_converge} discussed so far are calculated using $N_r=8$. This value is expected (and confirmed below) to be sufficient for the calculations where the HF monomers are in the ground vibrational state. However, since we are also interested in the vibrational states of the trimer for excited HF-stretch states, it is necessary to investigate their convergence with respect to the value of $N_r$. Performing the convergence tests for increasing $N_r$ values  by means of 12D calculations would be very demanding, since their computational cost scales as $(N_r)^3$. Therefore, the $N_r$-convergence of the $v=1,2$ HF stretching states of the trimer is assessed by diagonalizing the matrix of the 3D intramolecular Hamiltonian $\hat H_{\rm intra}$ for $N_r = 8, 10, 12, 16$. It is reasonable to assume that the value of $N_r$ which gives converged HF-stretch fundamentals and overtones in the 3D calculations will do the same in the 12D calculations as well. 

The results of such 3D calculations are presented in Table \ref{tab_3D_results}. It is evident from them that for the $v=1$ states $\nu^{\rm HF}_{\rm sym}$ and $\nu^{\rm HF}_{\rm asym}$ (a) the energies calculated with $N_r = 8$ differ by about -0.5 cm$^{-1}$ from those obtained for the larger $N_r$ values, and (b) using $N_r = 10$ gives results that differ by at most -0.05 cm$^{-1}$ from those calculated with $N_r = 12$ and 16. Similar observations hold for the four $v=2$ HF-stretch states. Their energies for $N_r=8$ differ by up to -7 cm$^{-1}$ from their counterparts obtained with $N_r = 10, 12, 16$. In contrast, the $N_r=10$ results agree with those for $N_r = 12$ and 16 to within -0.6 cm$^{-1}$.  The ``convergence from below'' behavior with respect to increasing $N_r$ implies that for the $v=1$ and 2 intramonomer excited states $N_r=8$ is not a dense enough grid for accurate quadrature over the $r$ coordinates. In view of this, the 12D results discussed in the following sections are those calculated using $N_r=10$, with other basis-set parameters being those of basis I. This basis set, with $N_r=10$, is designated as IA in Table \ref{tab_12D_bases} and elsewhere. 

It should be emphasized that for the intermolecular vibrational states of HF trimer listed in Table \ref{tab_12D_converge}, from 12D calculations and for the monomers in the ground intramolecular vibrational state, the energies computed using bases IA $(N_r=10)$ and I $(N_r=8)$ are virtually identical. For this reason, the second column of Table \ref{tab_12D_converge} is labeled as I/IA.

\subsection{Intermolecular vibrational eigenstates}\label{ss_intermolec}

Selected low-energy intermolecular vibrational states of HF trimer from full-dimensional 12D quantum calculations using basis set IA, for the monomers in their ground intramolecular vibrational states, and also the rigid-monomer 9D quantum calculations, are presented in Table \ref{tab_12D_states}. Inspection of the results shows that the 12D and 9D level energies typically differ by 1-4 cm$^{-1}$, but the differences can go up to 7 cm$^{-1}$. This demonstrates that the coupling between the intra- and intermolecular vibrational modes of the trimer is not negligible already for the ground states of the HF monomers. It has to be taken into account in accurate bound-state calculations and comparison with experiments.

The states in Table \ref{tab_12D_states} are assigned in terms of the fundamentals, overtones and combinations of the intermolecular symmetric and asymmetric stretch modes ($\nu_{ss}$ and $\nu_{as}$), respectively, as well as the following bending modes: a nondegenerate symmetric mode -- $\nu_{isb}$ (irrep $A_1'$) and $\nu_{osb}$ (irrep $A_1''$), for the in- and out-of-plane cases, respectively -- and two doubly-degenerate asymmetric modes -- $\nu_{iab}$ (irrep $E'$) and $\nu_{oab}$ (irrep $E''$) for the in- and out-of-plane cases, respectively. The assignments are made in the way described in I. It is evident that the fundamental frequencies of the stretching modes $\nu_{ss}$ and $\nu_{as}$, 186.9 and 170.9 cm$^{-1}$, respectively, are much lower than those of the bending modes, all of which are above 400 cm$^{1}$. Also shown for each 12D state is the basis-state norm (BSN), that measures the contribution of the dominant product inter/intra-basis state to the given eigenstate. For all 12D states shown the BSN is very close to 1, meaning that these eigenstates are highly pure, i.e., dominated by a single inter/intra-basis state.

Table \ref{12D_expect_values} shows for selected 12D eigenstates the expectation values $\langle r_k\rangle$, $\langle R_k\rangle$, and $\langle |\phi _k|\rangle$ of the coordinates defined in Sec. \ref{ssection:CHGA}, together with the corresponding root-mean-square (rms) amplitudes $\Delta r_k$, $\Delta R_k$, $\Delta |\phi _k|$, and $\Delta \theta_k$. These quantities are sensitive to the excitation of different intermolecular and intramolecular DOFs of the trimer. Thus, as expected, $\langle R_k\rangle$ (and $\Delta R_k$) increases with the number of quanta in the symmetric stretch mode, 4.953 bohrs for $\nu_{ss}$ and 5.007  bohrs for $2\nu_{ss}$, relative to the ground-state value of 4.901 bohrs. Interestingly, and somewhat unexpectedly, $\langle R_k\rangle$ values for the fundamentals of nondegenerate symmetric bending modes $\nu_{isb}$  and $\nu_{osb}$, 4.992 and 4.959 bohrs, respectively, are in fact larger than or comparable to that for the $\nu_{ss}$  symmetric-stretch fundamental, 4.953 bohrs. In all likelihood, this reflects the stretch-bend coupling present in the intermolecular modes, so that their ``stretch" and ``bend" designations need to be used with caution.

\subsection{Intramolecular vibrational eigenstates and frequency shifts}\label{ss_intramolec}

Table \ref{tab_12D_intra_2} gives the energies of the $v=1,2$ HF stretching states of HF trimer from 12D calculations, for intermolecular modes in the ground state, using basis IA with $N_r=10$. Also shown for comparison are the corresponding results from the 12D calculations with basis I, where $N_r=8$. One can see that for the $v=1$ states $\nu^{\rm HF}_{\rm sym}$ and $\nu^{\rm HF}_{\rm asym}$, the energies calculated using bases IA and I differ by less than 0.4 cm$^{-1}$. This agrees with the observation made earlier in Sec. \ref{ss_convergence}, in connection with the results of the reduced-dimension 3D calculations of HF trimer displayed in Table \ref{tab_3D_results}, that the  3D $v=1$ HF-stretch eigenstates calculated with bases IA and I differ by about 0.5 cm$^{-1}$. The conclusion is that basis set I, $N_r=8$, when used in 12D calculations, gives accurate energies of $v=1$ HF-stretch states of the trimer, as well as the intermolecular vibrational states in this intramolecular manifold (and of course, intermolecular vibrational states for the ground intramolecular state of HF monomers).

For the $v=2$ HF-stretch states of the trimer in Table \ref{tab_12D_intra_2}, their energies from the 12D calculation using bases IA and I differ by up to 4 cm$^{-1}$. Again, this is in line with the corresponding results for these states from 3D calculations reported in Table \ref{tab_3D_results}, which show that their energies for bases IA and I differ by up to 6.8 cm$^{-1}$.

Evidently, the trends in 12D results for bases IA and I mirror those from the 3D calculations for the two bases. As already discussed in Sec. \ref{ss_convergence} based on Table \ref{tab_3D_results}, the energies of $v=2$ HF-stretch states from 3D calculations with $N_r=10$ agree to within 0.6 cm$^{-1}$ with those obtained for $N_r=12$ and 16. This gives us confidence that the corresponding results of the 12D calculations using basis IA $(N_r=10)$ are converged to the same degree with respect to $N_r$.

Complex formation shifts the frequencies of the intramolecular vibrations of the constituent monomers away from the vibrational frequencies of the isolated monomers. Such vibrational frequency shifts were characterized accurately in our recent full-dimensional calculations of the inter- and intramolecular vibrational states of binary molecular complexes of H$_2$O/D$_2$O-CO,\cite{BACIC20C} HDO-CO,\cite{BACIC21} and HCl-H$_2$O.\cite{BACIC21A}  The 12D results in Table \ref{tab_12D_intra_2} allow us to calculate vibrational frequency shifts for HF trimer. On the PES employed, for the monomers at large separation, the $v=1$ and $v=2$ vibrational levels of the isolated HF are at 3959.83 and 7728.99 cm$^{-1}$, respectively. A glance at the 12D results in Table \ref{tab_12D_intra_2} shows that the frequencies of the $v=1,2$ HF stretching states of HF trimer are substantially redshifted in comparison to those of the isolated HF monomer. Thus, using the results for the IA basis, the frequency shifts (redshifts) of the states  $\nu^{\rm HF}_{\rm sym}$ and $\nu^{\rm HF}_{\rm asym}$ are -280.43 and -216.78 cm$^{-1}$, respectively. Based on the measured frequencies of the isolated HF-stretch fundamental,\cite{NIELSEN56} 3961 cm$^{-1}$ and that of the asymmetric H-F stretch $\nu^{\rm HF}_{\rm asym}$ fundamental,\cite{LISY86} 3712 cm$^{-1}$, the experimental value of the frequency shift of the $\nu^{\rm HF}_{\rm asym}$ fundamental of the trimer is -249 cm$^{-1}$, in good agreement with our theoretical result of -217 cm$^{-1}$. In Sec. \ref{ss_cooperative}, the calculated trimer redshifts are compared to that calculated for the HF dimer, in the context of cooperative hydrogen bonding.

\subsection{Effects of HF-stretch intramolecular excitation on the intermolecular vibrational states of HF trimer}\label{ss_intramolec}

The degree to which the intramolecular HF-stretch excitation affects the energies of the intermolecular vibrational states reflects the strength of the coupling between the intramolecular and intermolecular DOFs of the trimer. In order to gain insight into this intra/inter coupling, Table \ref{tab_inter_intra} lists the energies of the fundamentals of the intermolecular stretching and bending modes of the trimer in both the ground-state and excited $v=1$ $\nu^{\rm HF}_{\rm sym}$ and $\nu^{\rm HF}_{\rm asym}$ intramolecular vibrational manifolds, from 12D calculations using basis IA. It is immediately evident that the energies of all intermolecular vibrational modes considered increase significantly upon the excitation of either HF-stretch intramolecular state relative to those in the ground intramolecular vibrational state. In the HF-excited $\nu^{\rm HF}_{\rm sym}$ intramolecular manifold the energies of the stretching mode fundametals $\nu_{as}$ and $\nu_{ss}$ increase by 12.9 and 12.1 cm$^{-1}$, respectively, while for the fundamentals of the four bending modes the energy increases are larger, ranging from 26.5 cm$^{-1}$ $(\nu_{oab})$ to 34.9 cm$^{-1}$ $(\nu_{iab})$. Excitation of the $\nu^{\rm HF}_{\rm asym}$ intramolecular vibrational mode results in comparable increases in the energies of the intermolecular vibrational modes considered.

Additional evidence for the coupling between the intra- and intermolecular vibrational DOFs is provided by the two entries at the bottom of Table \ref{12D_expect_values} pertaining to the fundamental and the first overtone of the $\nu^{\rm HF}_{\rm sym}$ intramolecular mode, respectively. As expected, the vibrationally averaged HF bond length increases with the number of quanta in the $\nu^{\rm HF}_{\rm sym}$ mode, from 1.790 bohrs in the ground-state to 1.815 bohrs for the $\nu^{\rm HF}_{\rm sym}$ fundamental  and 1.841 bohrs for the $2\nu^{\rm HF}_{\rm sym}$ overtone. But what is interesting, and surprising, is that excitation of the $\nu^{\rm HF}_{\rm sym}$ mode also results in appreciable shortening of the (vibrationally averaged) distance between the HF monomers, from 4.901 bohrs in the ground state to 4.853  bohrs for the $\nu^{\rm HF}_{\rm sym}$ fundamental  and 4.815 bohrs for the $2\nu^{\rm HF}_{\rm sym}$ overtone. Clearly, the coupling between intra- and intermolecular modes of HF trimer manifests in the changes of both its vibrational energy level structure and geometric features due to the intramolecular vibrational excitations.

\subsection{Manifestations of cooperative hydrogen bonding}\label{ss_cooperative}

A distinguishing feature of HF trimer, and molecular trimers in general, relative to HF dimer is the cooperative hydrogen bonding resulting from the 3-body interactions which, of course, are absent in the dimer(s). The results of the 12D calculations in this paper reveal several manifestations of the cooperativity in the hydrogen bonding of HF trimer. One of them pertains to the vibrational frequency shifts of the HF-stretching states of HF trimer, relative to the stretch fundamental of the isolated HF monomer. In Sec. \ref{ss_intramolec}, we stated that based on the 12D calculations, the frequency shifts (redshifts) of the symmetric and asymmetric H-F stretch fundamentals of the trimer,  $\nu^{\rm HF}_{\rm sym}$ and $\nu^{\rm HF}_{\rm asym}$, are -280.43 and -216.78 cm$^{-1}$, respectively. It is instructive to compare these two trimer redshifts to the bound-HF $(\nu_2$) stretching fundamentals of HF dimer, calculated in 6D\cite{AVOIRD03} on the SO-3 PES to be -92.74 cm$^{-1}$. Evidently, the calculated redshifts of the symmetric and asymmetric H-F stretch fundamentals of HF trimer are much larger by magnitude than the redshift of the stretching fundamental of the hydrogen-bond donor (bound) HF in HF dimer. This can only be due to the cooperative hydrogen bonding and the 3-body interactions present in HF trimer, but not in HF dimer.

Another manifestation of the cooperative hydrogen bonding in HF trimer involves the separation of the monomers in the trimer. From Table \ref{12D_expect_values}, the 12D vibrationally averaged ground-state distance between the centers of mass of the monomers in HF trimer is 4.901 bohrs. It is significantly shorter than the vibrationally averaged separation of the monomers in the ground state of HF dimer, 5.24 bohrs, from the 6D calculations on the SO-3 PES.\cite{AVOIRD03} This decrease in the intermonomer distance in HF trimer, relative to the dimer, can also be attributed to the cooperative hydrogen bonding, i.e., the 3-body interactions. Similar shortening of the intermonomer distance with increasing cluster size has been reported for H$_2$O clusters,\cite{SAYKAL963} and interpreted in terms of hydrogen-bond cooperativity.

Cooperative hydrogen bonding manifests in the energetics of HF trimer as well. The 12D binding energy of HF trimer relative to the energy of the three separated HF monomers in the ground state is 3662.35 cm$^{-1}$ (footnote of Table \ref{tab_12D_states}). The binding energy of the isolated HF dimer is 1061.73 cm$^{-1}$, from the 6D calculations on the SO-3 PES.\cite{AVOIRD03} In the absence of 3-body interactions, the trimer binding energy would be simply three times that of the isolated HF dimer, i.e., $3\times 1061.73 = 3185.19$ cm$^{-1}$. But this is 477.16 cm$^{-1}$ less than the actual 12D-calculated binding energy of the trimer, 3662.35 cm$^{-1}$, evidence that the 3-body interactions contribute to, and increase significantly, the binding energy of the trimer.

The binding energy of HF trimer with respect to the HF$_2$ (g.s.) + HF channel can also be readily calculated. Since the 12D binding energy of HF trimer relative to that of the three HF (g.s.) monomers is 3662.35 cm$^{-1}$, and the 6D binding energy of the isolated HF dimer is 1061.73 cm$^{-1}$,\cite{AVOIRD03} the binding energy of HF trimer relative to HF$_2$ (g.s.) + HF is $3662.35 - 1061.73 = 2600.62$ cm$^{-1}$.

\subsection{Comparison between theory and experiment}\label{ss_comparison}

Spectroscopic data regarding the vibrational frequencies of HF trimer are limited. The available information from spectroscopic measurements in the gas phase and Ne and Ar matrices is presented Table \ref{tab_calc_obs}, together with the corresponding theoretical results from the 12D calculations. Closer inspection shows that the calculated fundamental frequencies of the intermolecular $\nu_{as}$,  $\nu_{iab}$,  and $\nu_{osb}$ modes are 1.4-2.7\% higher than the respective experimental values measured in Ne matrix ($\nu_{as}$) and the gas phase ($\nu_{iab}$ and $\nu_{osb}$). As for the intramolecular vibrational fundamentals, only one (gas phase) experimental piece of information is available, 3712 cm$^{-1}$, for the $\nu^{\rm HF}_{\rm asym}$ mode. The corresponding calculated (in 12D) value is 3743 cm$^{-1}$, 0.8\% larger than the measured value. Therefore, at this point the overall agreement between theory and the scant experimental data can be described as reasonable, but not excellent. Since the quantum calculations are full-dimensional and well-converged, the discrepancies between theory and experiment can be attributed to the shortcomings of the PES employed.

\section{Conclusions}\label{s_conclusions}

In this paper we present the methodology for full-dimensional (12D) quantum calculations of the coupled intramolecular and intermolecular vibrational states of noncovalently bound trimers of flexible diatomic molecules. It builds on our recently introduced approach for fully coupled 9D quantum calculations of the vibrational states of trimers comprised of diatomics treated as rigid,\cite{BACIC22B} and extends it to include in a rigorous manner the intramolecular stretching coordinates of the three diatomic monomers. The coordinates and the 12D vibrational Hamiltonian for HF trimer developed by Wang and Carrington\cite{CARRING01B} are employed. At the heart of our methodology is the use of the eigenstates of reduced-dimension Hamiltonians as contracted basis functions for both intermolecular and intramolecular DOFs of the trimer. This computational strategy introduced by us\cite{BACIC19B,BACIC20C} has proven to be very successful in full-dimensional quantum calculations of (ro)vibrational states of a variety of binary molecular complexes.\cite{BACIC22C,CARRING23} Here, a 9D rigid-monomer intermolecular vibrational Hamiltonian of the trimer is diagonalized as done previously,\cite{BACIC22B} in a product contracted basis of the eigenvectors of a 3D ``frame'' (intermolecular stretching) Hamiltonian and a  6D ``bend'' Hamiltonian.  In addition, a 3D intramolecular vibrational Hamiltonian for the three monomers is diagonalized separately. A certain number of the lower-energy 9D intermolecular and 3D intramolecular vibrational eigenstates is included in the 12D product contracted basis in which the full vibrational Hamiltonian of the trimer is diagonalized. Thus, what makes these 12D calculations feasible is a sequence of diagonalizations of lower-dimensional Hamiltonians, each step generating a contracted  eigenstate basis for  diagonalizing a  Hamiltonian of higher dimensionality in the next step. Heavy use is made of the  $G_{12}$ symmetry of (HF)$_3$ to block-diagonalize the matrix of the full 12D Hamiltonian into the blocks associated with the $G_{12}$ irreps, by constructing symmetry-adapted product-contracted basis functions. As in Ref. \onlinecite{BACIC22B}, the 2+3-body SO-3 + HF3BG PES of HF trimer\cite{QUACK98,QUACK01} is employed in the calculations reported in this paper.

The present rigorous 12D quantum calculations of the fully coupled intra- and intermolecular vibrational states of HF trimer are the first such calculations for a hydrogen-bonded trimer of diatomic molecules treated as flexible. From the  earlier 9D rigid monomer calculations\cite{BACIC22B} we concluded that the intermolecular vibrations exhibit strong stretch-bend coupling, and that the nonadditive three-body interactions make a large contribution to the (HF)$_3$ PES and affect the intermolecular vibrational energy levels. These conclusions remain unchanged in the 12D calculations. The novel aspects of the vibrational quantum dynamics of the trimer that the present 12D calculations elucidate are those associated with the intramolecular HF-stretch vibrational coordinates of the three monomers and their coupling with the intermolecular DOFs of the trimer. They are beyond the scope of the rigid-monomer treatment. Thus, the energies of the 12D intermolecular vibrational states of the trimer differ by those from the 9D rigid-monomer calculations typically by 1-4 cm$^{-1}$, and by as much as 7 cm$^{-1}$, already when the monomers are in their ground intramolecular vibrational state. This points to the non-negligible coupling between the intra- and intermolecular vibrational DOFs of the trimer even when the monomers are not vibrationally excited. Also reported are the energies of the $v=1,2$ HF-stretch excited intramolecular vibrational states, together with the intermolecular vibrational states in these intramolecular manifolds. These results provide additional evidence for substantial coupling between the intramolecular and intermolecular DOFs of the trimer. Excitation of either of the two $v=1$ HF-stretch intramolecular states causes significant increase in the energies of all intermolecular vibrational modes considered relative to their values in the ground intramolecular vibrational state.  In another manifestation of the coupling between the inter- and intramolecular DOFs,  one- and two-quanta excitations of the $\nu^{\rm HF}_{\rm sym}$ mode results in appreciable shortening of the (vibrationally averaged) distance between the HF monomers.

Complexation results in the shift of the frequencies of the intramolecular vibrations of the monomers forming (HF)$_3$  away from the vibrational frequencies of the isolated monomers. The 12D calculations show that the frequencies of the $v=1,2$ HF stretching states of HF trimer are strongly redshifted in comparison to those of the isolated HF monomer, in agreement with spectroscopic data. Moreover, the magnitudes of these trimer redshifts are much larger than that of the redshift of the stretching fundamental of the hydrogen-bond donor (bound) HF in HF dimer. The most likely explanation for this is the cooperative hydrogen bonding in HF trimer, arising from the 3-body interactions, which do not exist in HF dimer. The 12D calculations also reveal manifestations of cooperative hydrogen bonding in the structural properties and the energetics of HF trimer.

Comparison is made between the 12D theoretical results and the limited experimental information from the spectroscopic measurements of (HF)$_3$ in the gas phase and Ne and Ar matrices. The fundamental frequencies of the intermolecular $\nu_{as}$,  $\nu_{iab}$,  and $\nu_{osb}$ modes from the 12D calculations are 1.4-2.7\% higher than the respective experimental values from the spectroscopic measurements in Ne matrix and the gas phase. An experimental value  is available for only one intramolecular vibrational fundamental, that of the $\nu^{\rm HF}_{\rm asym}$ mode. It can be compared to the 12D result which is 0.8\% larger. Clearly, there is room for improvement in the agreement between theory and experiment regarding the intra- and intermolecular vibrations of HF trimer. Given that the 12D quantum calculations are full-dimensional and well-converged, what remains is to develop an improved 12D PES of (HF)$_3$. More comprehensive spectroscopic data pertaining to the intra- and intermolecular vibrational excitations of this trimer  are needed as well.

The 12D methodology presented here can, and will, be applied to other similar trimers, e.g., HCl trimer. It can also provide a starting point for the rigorous quantum treatment of different but related molecular trimers such as halide dihydrates. It is reasonable to expect that this approach will soon enable fully coupled 12D quantum calculations of the vibration-rotation-tunneling states of H$_2$O trimer in the rigid-monomer approximation. They would allow a comprehensive interpretation of the remarkable measured low-frequency spectra of water trimer in helium nanodroplets.\cite{BOWMAN20}

\section*{Supplementary Material}

The procedure for computing the $\hat H_{\rm inter}$ eigenstates of $E_b'/E_b''$ symmetry directly from the $E_a'/E_a''$ eigenstates of $\hat H_{\rm inter}$ is given in Sec. I of the supplementary material. The full exploitation of permutation symmetry in computing the 12D matrix elements of $\Delta V$ is presented in Sec. II of the supplementary material. The calculation of 12D expectation values of geometric quantities is described in Sec. III of the supplementary material.

\section*{Acknowledgments}
 
Z.B. and P.M.F. are grateful to the National Science Foundation for its partial support of this research through the Grants CHE-2054616 and CHE-2054604, respectively. P.M.F. is grateful to Prof. Daniel Neuhauser for his support. 

\section*{Author Declarations}
\noindent
{\bf Conflict of interest}

The authors have no conflicts to disclose.

\bigskip
\noindent
{\bf Author Contributions}

\medskip
\noindent 
{\bf Peter M. Felker:} Conceptualization (equal); Formal analysis (equal); Funding acquisition (equal); Methodology (equal); Writing - original draft (equal); Writing - review \& editing (equal). {\bf Zlatko Ba\v{c}i\'{c}:} Conceptualization (equal); Formal analysis (equal); Funding acquisition (equal); Methodology (equal); Writing - original draft (equal); Writing - review \& editing (equal). 

\section*{Data Availability}
 
 The data that support the findings of this study are available within the article and its supplementary material.

\newpage

%\bibliography{bacic}

\begin{thebibliography}{10}

\bibitem{BACIC96C}
Z.~Ba\v{c}i\'{c} and R.~E. Miller, J. Phys. Chem. {\bf 100}, 12945 (1996).

\bibitem{AVOIRD00A}
P.~E.~S. Wormer and A.~van~der Avoird, Chem. Rev. {\bf 100}, 4109 (2000).

\bibitem{CARRING11A}
Jr. T.~Carrington and X.-G. Wang, Wiley Interdiscip. Rev. Comput. Mol. Sci.
  {\bf 1}, 952 (2011).

\bibitem{AVOIRD22}
A.~van~der Avoird.
\newblock {Vibration-rotation-tunneling levels and spectra of Van der Waals
  molecules}.
\newblock In {\em Vibrational dynamics of molecules}, edited by J.~M. Bowman,
  page 194.  (World Scientific, Singapore, 2022).

\bibitem{MATYUS23}
E.~M\'{a}tyus, A.~Mart\'{i}n~Santa Daria and G.~Avila, Chem. Commun. {\bf 59},
  366 (2023).

\bibitem{BACIC95}
D.~H. Zhang, Q.~Wu, J.~Z.~H. Zhang, M.~von Dirke and Z.~Ba\v{c}i\'{c}, J. Chem.
  Phys. {\bf 102}, 2315 (1995).

\bibitem{BACIC98A}
Z.~Ba\v{c}i\'{c} and Y.~Qiu.
\newblock {Vibration-rotation-tunneling dynamics of (HF)$_2$ and (HCl)$_2$ from
  full-dimensional quantum bound-state calculations}.
\newblock In {\em Advances in Molecular Vibrations and Collision Dynamics, Vol.
  3}, edited by J.~M. Bowman and Z.~Ba\v{c}i\'{c}, page 183.  (JAI Press Inc.,
  Stamford, 1998).

\bibitem{BACIC97A}
Y.~Qiu and Z.~Ba\v{c}i\'{c}, J. Chem. Phys. {\bf 106}, 2158 (1997).

\bibitem{WU95}
Q.~Wu, D.~H. Zhang and J.~Z.~H. Zhang, J. Chem. Phys. {\bf 103}, 2548 (1995).

\bibitem{AVOIRD03}
G.~W.~M. Vissers, G.~C. Groenenboom and A.~van~der Avoird, J. Chem. Phys. {\bf
  119}, 277 (2003).

\bibitem{BACIC981}
Y.~Qiu, J.~Z.~H. Zhang and Z.~Ba\v{c}i\'{c}, J. Chem. Phys. {\bf 108}, 4804
  (1998).

\bibitem{HUANG19}
J.~Huang, D.~Yang, Y.~Zhou and D.~Xie, J. Chem. Phys. {\bf 150}, 154302 (2019).

\bibitem{CARRING18}
X.~{-G}. Wang and T.~{Carrington}, J. Chem. Phys. {\bf 148}, 074108 (2018).

\bibitem{BACIC19B}
P.~M. Felker and Z.~Ba\v{c}i\'{c}, J. Chem. Phys. {\bf 151}, 024305 (2019).

\bibitem{BACIC22C}
P.~M. Felker and Z.~Ba\v{c}i\'{c}, Phys. Chem. Chem. Phys. {\bf 24}, 24655
  (2022).

\bibitem{BACIC19A}
D.~Lauvergnat, P.~M. Felker, Y.~Scribano, D.~M. Benoit and Z.~Ba\v{c}i\'{c}, J.
  Chem. Phys. {\bf 150}, 154303 (2019).

\bibitem{BACIC20C}
P.~M. Felker and Z.~Ba\v{c}i\'{c}, J. Chem. Phys. {\bf 153}, 074107 (2020).

\bibitem{BACIC86}
Z.~Ba\v{c}i\'{c} and J.~C. Light, J. Chem. Phys. {\bf 85}, 4594 (1986).

\bibitem{BACIC87}
Z.~Ba\v{c}i\'{c} and J.~C. Light, J. Chem. Phys. {\bf 86}, 3065 (1987).

\bibitem{BACIC88}
Z.~Ba\v{c}i\'{c}, R.~M. Whitnell, D.~Brown and J.~C. Light, Comput. Phys.
  Commun. {\bf 51}, 35 (1988).

\bibitem{BACIC89}
Z.~Ba\v{c}i\'{c} and J.~C. Light, Annu. Rev. Phys. Chem. {\bf 40}, 469 (1989).

\bibitem{CARRING02}
X.~{-G}. Wang and T.~{Carrington, Jr.}, J. Chem. Phys. {\bf 117}, 6923 (2002).

\bibitem{CARRING03}
X.~{-G}. Wang and T.~{Carrington, Jr.}, J. Chem. Phys. {\bf 119}, 101 (2003).

\bibitem{CARRING06}
Jean~Christophe Tremblay and T.~{Carrington}, J. Chem. Phys. {\bf 125}, 094311
  (2006).

\bibitem{CARRING08}
X.~{-G}. Wang and T.~{Carrington, Jr.}, J. Chem. Phys. {\bf 129}, 234102
  (2009).

\bibitem{BOWMAN03A}
S.~Zhou, J.~M. Bowman and A.~Brown, J. Chem. Phys. {\bf 118}, 10012 (2003).

\bibitem{BACIC21}
P.~M. Felker and Z.~Ba\v{c}i\'{c}, J. Phys. Chem. A {\bf 125}, 980 (2021).

\bibitem{BACIC21A}
Y.~Liu, J.~Li, P.~M. Felker and Z.~Ba\v{c}i\'{c}, Phys. Chem. Chem. Phys. {\bf
  23}, 7101 (2021).

\bibitem{BACIC21B}
P.~M. Felker, Y.~Liu, J.~Li and Z.~Ba\v{c}i\'{c}, J. Phys. Chem. A {\bf 125},
  6437 (2021).

\bibitem{BACIC21C}
P.~M. Felker and Z.~Ba\v{c}i\'{c}, Chin. J. Chem. Phys. {\bf 34}, 728 (2021).

\bibitem{BACIC19D}
P.~M. Felker, D.~Lauvergnat, Y.~Scribano, D.~M. Benoit and Z.~Ba\v{c}i\'{c}, J.
  Chem. Phys. {\bf 151}, 124311 (2019).

\bibitem{BACIC20}
P.~M. Felker and Z.~Ba\v{c}i\'{c}, J. Chem. Phys. {\bf 152}, 014108 (2020).

\bibitem{BACIC20A}
P.~M. Felker and Z.~Ba\v{c}i\'{c}, J. Chem. Phys. {\bf 152}, 124103 (2020).

\bibitem{CARRING23}
X.~{-G}. Wang and T.~Carrington, J. Chem. Phys. {\bf 158}, 084107 (2023).

\bibitem{BACIC22B}
P.~M. Felker and Z.~Ba\v{c}i\'{c}, J. Chem. Phys. {\bf 157}, 194103 (2022).

\bibitem{CLARY02}
L.~S. Costa and D.~C. Clary, J. Chem. Phys. {\bf 117}, 7512 (2002).

\bibitem{YU04}
H.~{-G}. Yu, J. Chem. Phys. {\bf 120}, 2270 (2004).

\bibitem{LISY88}
K.~D. Kolebrander, C.~E. Dykstra and J.~M. Lisy, J. Chem. Phys. {\bf 88}, 5995
  (1988).

\bibitem{CARRING01}
X.~{-G}. Wang and T.~{Carrington, Jr.}, J. Chem. Phys. {\bf 115}, 9781 (2001).

\bibitem{CARRING01B}
X.~{-G}. Wang and T.~{Carrington, Jr.}, Can. J. Phys. {\bf 79}, 623 (2001).

\bibitem{QUACK93}
M.~Quack, J.~Stohner and M.~A. Suhm, J. Mol. Struct. {\bf 294}, 33 (1993).

\bibitem{QUACK01}
M.~Quack, J.~Stohner and M.~A. Suhm, J. Mol. Struct. {\bf 599}, 381 (2001).

\bibitem{QUACK98A}
M.~Quack and M.~A. Suhm.
\newblock {Spectroscopy and quantum dynamics of hydrogen fluoride clusters}.
\newblock In {\em Advances in Molecular Vibrations and Collision Dynamics, Vol.
  3}, edited by J.~M. Bowman and Z.~Ba\v{c}i\'{c}, page 205.  (JAI Press Inc.,
  Stamford, 1998).

\bibitem{QUACK98}
W.~Klopper, M.~Quack and M.~A. Suhm, J. Chem. Phys. {\bf 108}, 10096 (1998).

\bibitem{NESBITT93}
M.~A. Suhm, J.~T. Farrell, S.~H. Ashworth and D.~J. Nesbitt, J. Chem. Phys.
  {\bf 98}, 5985 (1993).

\bibitem{ROYP14}
P.~Asselin, P.~Soulard, B.~Madebene, M.~Goubet, T.~R. Huet, R.~Georges,
  O.~Pirali and P.~Roy, Phys. Chem. Chem. Phys. {\bf 16}, 4797 (2014).

\bibitem{LISY86}
D.~W. Michael and J.~M. Lisy, J. Chem. Phys. {\bf 85}, 2528 (1986).

\bibitem{ANDREWS84}
L.~Andrews, V.~E. Bondybey and J.~H. English, J. Chem. Phys. {\bf 81}, 3452
  (1984).

\bibitem{ANDREWS92}
L.~Andrews, S.~R. Davis and R.~D. Hunt, Mol. Phys. {\bf 77}, 993 (1992).

\bibitem{ANDREWS99}
L.~Andrews and P.~F. Souter, J. Chem. Phys. {\bf 111}, 5995 (1999).

\bibitem{MANDELSTAMTAYLOR97}
V.~A. Mandelshtam and H.~S. Taylor, J. Chem. Phys. {\bf 106}, 5085 (1997).

\bibitem{NEUHAUSER95}
M.~R. Wall and D.~Neuhauser, J. Chem. Phys. {\bf 102}, 8011 (1995).

\bibitem{CARRING94}
M.~J. Bramley and T.~{Carrington, Jr.}, J. Chem. Phys. {\bf 101}, 8494 (1994).

\bibitem{NIELSEN56}
G.~A. Kuipers, D.F. Smith and A.~H. Nielsen, J. Chem. Phys {\bf 25}, 275
  (1956).

\bibitem{SAYKAL963}
K.~Liu, J.~D. Cruzan and R.~J. Saykally, Science {\bf 271}, 929 (1996).

\bibitem{BOWMAN20}
R.~Schwan, C.~Qu, D.~Mani, N.~Pal, G.~Schwaab, J.~M. Bowman, G.~Tschumper and
  M.~Havenith, Angew. Chem. Int. Ed. {\bf 59}, 11399 (2020).

\end{thebibliography}

\clearpage

\begin{table}[ht]
\caption{Compositions of the various 12D basis sets.}
\centering
\begin{tabular}{ccccccccccc}
\hline\hline
Basis & $\bar r$/bohrs & $N_r$ & $N_R$ & $N_F$ & $N_B$ & $N_{\rm inter}$ & $N_{intra}$ & $N_{A}$ & $N_{E}$ \\ \hline
I& 1.7950 &8& 12 & 201 & 120 & 600 & 56 & 5700 & 11100 \\
IA& 1.7950 &10& 12 & 201 & 120 & 600 & 56 & 5700 & 11100 \\
II& 1.7950 &8& 12 & 201 & 120 & 540 & 56 & 5100 & 10020 \\
III & 1.7950 &8& 14 & 201 & 120 & 480 & 56 & 4560 & 8880 \\
IV & 1.7950 &8& 12 & 150 & 120& 396 & 56 & 3762 & 7326 \\
V & 1.7843 &8& 12 & 201 & 120 & 600 & 56 & 5700 & 11100 \\
VI & 1.7843 &8& 12 & 150 & 120 & 240 & 56 & 2280 & 4440 \\
VII & 1.7813 &8& 12 & 150 & 120 & 280 & 56 & 2660 & 5180 \\
\hline

\multicolumn{10}{l}{The $N_r$ values pertain to the computation of the $\hat H_{\rm intra}$ eigenstates included in the 12D basis.}\\ 
\multicolumn{10}{l}{The $N_R$ values pertain to the computation of the frame states included in the 9D basis.}\\ 
\multicolumn{10}{l}{The $\bar r$ values are those used to produce the frame and bend states included in the 9D basis.}\\ 
\multicolumn{10}{l}{$N_F$ is the total number of 3D frame states of all symmetries used to build the 9D basis.}\\
\multicolumn{10}{l}{$N_B$ is the total number of 6D bend states of a given parity used to build the 9D basis.}\\
\multicolumn{10}{l}{ $N_{\rm inter}$ is the total number of 9D intermolecular states of a given parity used to build the 12D basis.}\\
\multicolumn{10}{l}{ $N_{\rm intra}$ is the total number of 3D intramolecular states used to build the 12D basis.}\\
\multicolumn{10}{l}{ $N_{A}$ and $N_{E}$ are the dimensions of the $A$ and $E$ blocks of the 12D $\hat H$ matrices.}\\
 
\end{tabular}
\label{tab_12D_bases}
\end{table}

 \clearpage
 
 \begin{table}[ht]
\caption{Basis-set dependence of the energies$^a$ of the $v=1,2$ HF stretching states in HF trimer from 12D calculations, for the intermolecular modes in the ground state. $N_r = 8$  in basis sets I-VII. All energies are in cm$^{-1}$.}
\centering
\begin{tabular}{cccccccc}
\hline\hline
 I & II & III & IV & V & VI & VII  & Assign.\\ \hline
 3679.04 & 3679.10 & 3679.17 & 3679.19 &  3679.52 & 3680.39 & 3680.50 & $\nu^{\rm HF}_{\rm sym}$  \\
 3742.69 & 3742.76 & 3742.77 & 3742.81 &  3743.06 & 3743.78 &  3743.82 &$\nu^{\rm HF}_{\rm asym}$  \\
7201.27 & 7201.65 & 7201.81 & 7201.97& 7202.72 & 7206.42 &7206.54 &  $2\nu^{\rm HF}_{\rm sym}$   \\ 
 7221.25&  &  &  & 7222.59 &  &  & $ \nu^{\rm HF}_{\rm sym}+ \nu^{\rm HF}_{\rm asym}$ \\
 \{7397.49\}$^b$&  &  &  & \{7399.38\}$^b$ &  &  & $2\nu^{\rm HF}_{\rm asym}(A_1')$ \\
 7452.90 &  &  &  &  7454.13 &  &  & $2\nu^{\rm HF}_{\rm asym}(E')$ \\
 \hline
\hline
\multicolumn{8}{l}{$^a$ The listed energies are all relative to the relevant ground-state energy.}\\
\multicolumn{8}{l}{$^b$ Significant basis-state mixing leading to small BSN.}\\

\end{tabular}
\label{tab_12D_intra}
\end{table}

\clearpage

\begin{table}[ht]
\caption{Basis-set dependence of the energies$^a$ of selected low-energy intermolecular vibrational eigenstates of HF trimer from 12D calculations, for the HF monomers in their ground intramolecular vibrational state. All energies are in cm$^{-1}$.}
\centering
\begin{tabular}{ccccccccc}
\hline\hline
Basis & I/IA$^b$ & II & III & IV & V & VI & VII & Assign. \\ \hline
$A_1',A_2'$\\
1 & 0 & 0 & 0 & 0& 0& 0& 0&g.s. \\
2 & 186.90 & 187.03 & 187.09 & 187.11 & 187.00 & 188.00 & 187.84 & $\nu_{ss}$ \\
3 & 328.24 &328.34 & 328.39 & 328.66 & 328.24 & 329.32 & 328.73 & $2\nu_{as}$ \\
4 & 368.92 &369.66 & 369.64 & 369.76 & 369.37 & 371.37 &  371.31 &$2\nu_{ss}$ \\
18 & 776.71 &776.64 & 776.71 & 776.67 & 776.10 & 775.68 &775.66 &$\nu_{isb}$ \\
\hline
$E'$\\
1,2 & 170.91 &170.93 & 171.23 & 171.46 &  171.04 & 173.50 & 172.17 &$\nu_{as}$ \\
3,4 & 332.73 &332.84 & 334.04 & 335.35 &  332.93 & 334.78 & 334.74 &$2\nu_{as}$ \\
5,6 & 351.43 &351.52 & 352.74 & 353.49 &  351.92 & 353.07 & 353.08 &$\nu_{as} + \nu_{ss}$ \\
9,10 & 501.44 &501.42 &501.54 &  501.73 &  501.25 & 502.01 & 501.86 &$\nu_{iab}$ \\
\hline
$A_1'',A_2''$ \\
1 & 556.17 & 556.26 &  & & 556.30 & & &$\nu_{as}+\nu_{oab}$ \\
2 & 583.98 & 584.06 &  &  & 584.09 & & &$\nu_{as}+\nu_{oab}$ \\
3 & 617.99 &618.01 &  & & 618.14 & & &$\nu_{osb}$\\ \hline
$E_a'',E_b''$ \\
1,2 & 416.77 & 416.76 & & & 417.06 & & &$\nu_{oab}$ \\
3,4 & 576.47 &576.50 & & & 576.33 & & &$\nu_{as}+\nu_{oab}$ \\
5,6 & 599.23 &  599.22   & & & 599.10 & & & \\
7,8 & 699.77 & 699.86  & & & 699.39 & & & \\

\hline
\hline
\multicolumn{9}{l}{$^a$ The listed energies are all relative to the relevant ground-state energy.}\\
\multicolumn{9}{l}{$^b$ The listed energies are from basis IA. The basis-I energies differ from the IA ones by 0.02 cm$^{-1}$ or less.}\\
\multicolumn{9}{l}{The $D_0$ values corresponding to the ground states of bases I/IA to VII are, respectively,}\\ 
\multicolumn{9}{l}{3662.37/3662.35, 3662.35, 3662.28, 3662.25, 3662.18, 3661.31, and 3661.30 cm$^{-1}$. These compare to the}\\ 
\multicolumn{9}{l}{$D_e$ value of $-5391.33$ cm$^{-1}$ for the complete dissociation of the trimer on the SO-3 + HF3BG PES.\cite{CARRING01}}\\ 

\end{tabular}
\label{tab_12D_converge}
\end{table}

\clearpage

\begin{table}[ht]
\caption{Low-energy eigenstates of $\hat H_{\rm intra}$ for four different values of $N_r$, from 3D calculations. The $\Delta E(N_r)$ values are eigenenergies (cm$^{-1}$) relative to the relevant ground-state energy.$^a$}
\centering
\begin{tabular}{cccccc}
\hline\hline
Assign.  & irrep &  $\Delta E(8)$ & $\Delta E(10)$ & $\Delta E(12)$ & $\Delta E(16)$ \\ \hline
  g.s. &$A_1^\prime$ & 0.00 &       0.00 & 0.00 & 0.00  \\
  $\nu^{\rm HF}_{\rm sym}$ & $A_1'$ & 3733.36 & 3733.90 & 3733.94 &    3733.95       \\
  $\nu^{\rm HF}_{\rm asym}$ & $E'$ & 3784.82 & 3785.34 & 3785.38 &    3785.39      \\
   $2\nu^{\rm HF}_{\rm sym}$ & $A_1'$ & 7325.44 & 7331.92 & 7332.43 &   7332.48      \\
  $\nu^{\rm HF}_{\rm sym}+\nu^{\rm HF}_{\rm asym}$ & $E'$ & 7337.21 & 7344.03 & 7344.57 &    7344.62       \\
  $2\nu^{\rm HF}_{\rm asym}$ & $A_1'$ & 7510.38 & 7511.83 & 7511.94  &  7511.96      \\
    $2\nu^{\rm HF}_{\rm asym}$ & $E'$ & 7550.57 & 7551.86 & 7551.75  &  7551.77      \\
\hline

\multicolumn{6}{l}{$^a$ The ground-state energies for $N_r=8$, 10, 12, and 16 are, respectively,}\\
\multicolumn{6}{l}{880.88, 880.95, 880.96, and 880.96 cm$^{-1}$ relative to the }\\ 
\multicolumn{6}{l}{potential energy of the three isolated monomers in their equilibrium geometries.}\\
\end{tabular}
\label{tab_3D_results}
\end{table}

 \clearpage
 
 \begin{table}[ht]
\caption{Low-energy intermolecular vibrational states of HF trimer from 12D and 9D (rigid-monomer) calculations, basis set IA (in cm$^{-1}$). The 12D results are for the HF monomers in their ground intramolecular vibrational state. }
\centering
\begin{tabular}{ccccc}
\hline\hline
&$\Delta E$ (12D) & $\Delta E$ (9D) & BSN$^c$ & Assignment\\ \hline
$A_1',A_2'$\\
1& 0.0$^a$  & 0.0$^b$ & 0.996 & g.s. \\
2 & 186.90 & 190.77 & 0.995 & $\nu_{ss}$ \\
3 & 328.24 & 329.51 & 0.996 & $ 2\nu_{as}$ \\
4 & 368.92 & 376.08 & 0.993 & $2\nu_{ss}$ \\
5 & 476.00 & 479.90 & 0.992 &     \\
18 & 776.71 & 775.96 & 0.896 & $\nu_{isb}$ \\  \hline
$E'$\\
1,2 & 170.91 & 171.09 & 0.996 & $\nu_{as}$ \\
3,4 & 332.73 & 333.96 & 0.992 & $2\nu_{as}$ \\
5,6 & 351.43 & 353.97 & 0.991 & $\nu_{as} + \nu_{ss}$ \\
9,10 & 501.44 & 498.29 & 0.994 & $\nu_{iab}$ \\ \hline
$A_1'',A_2''$ \\
1 & 556.17 & 554.98 & 0.996 & $\nu_{as}+ \nu_{oab}$  \\
2 & 583.98 & 582.26 & 0.996  & $\nu_{as} + \nu_{oab}$  \\
3 & 617.99 & 614.25 & 0.996 & $\nu_{osb}$  \\ 
4 & 707.98 & 709.14 & 0.993 &    \\ \hline
$E''$ \\
1,2 & 416.77 & 414.32 & 0.996  & $\nu_{oab}$  \\
3,4 & 576.47 & 573.27 & 0.994 & $\nu_{as} +\nu_{oab}$  \\
5,6 & 599.23 & 597.69 & 0.993  &  \\
7,8 &699.77 & 698.27 & 0.995 &  \\ 

 \hline
\hline
\multicolumn{5}{l}{$^a$The 12D ground-state energy is -3662.35 cm$^{-1}$ relative to the energy of separated flexible monomers.}\\
 \multicolumn{5}{l}{$^b$The 9D ground-state energy is -3812.81 cm$^{-1}$ relative to the energy of the separated rigid monomers.}\\
\multicolumn{5}{l}{$^c$ Basis-state norm of dominant basis state. It refers to the 12D results. See text for definition.}\\

\end{tabular}
\label{tab_12D_states}
\end{table}

\clearpage

\begin{table}[ht]
\caption{Expectation values of some coordinates of HF trimer for selected 12D eigenstates. $\langle r_k\rangle$, $\Delta r_k$, $\langle R_k\rangle$, and $\Delta R_k$ are in bohrs, while $\langle |\phi _k|\rangle$, $\Delta |\phi _k|$, and $\Delta \theta_k$ are in degrees. See text for definitions.}
\centering
\begin{tabular}{ccccc}
\hline\hline
 State & $\langle r_k\rangle$ ($\Delta r_k$) & $\langle R_k\rangle$ ($\Delta R_k$) & $\langle |\phi _k|\rangle $ ($\Delta |\phi _k|$) & $\Delta \theta_k^a$ \\ \hline
g.s. & 1.790 (0.129) &  4.901 (0.181) & 56.2 (10.9) & 11.9   \\
$\nu_{ss}$ & 1.789 (0.129) & 4.953 (0.260) & 56.5 (11.3) & 12.1 \\
$2\nu_{ss}$ & 1.787 (0.129) & 5.007 (0.319) & 56.8 (11.6) & 12.3 \\
$\nu_{isb}$ & 1.786 (0.128) & 4.992 (0.204) & 58.4 (15.3) & 14.8 \\
$\nu_{osb}$ & 1.787 (0.129) & 4.959 (0.196) & 57.1 (11.8) & 15.4 \\
$\nu^{\rm HF}_{\rm sym}$ & 1.815 (0.171) & 4.853 (0.176) & 55.4 (10.6) & 11.5  \\
$2\nu^{\rm HF}_{\rm sym}$ & 1.841 (0.212) & 4.815 (0.171) & 54.9 (10.4)  & 11.2   \\ 
 \hline
\hline
\multicolumn{5}{l}{$^a$ The values of $\langle \theta _k\rangle$ are all 90$^\circ$ by symmetry.}\\

\end{tabular}
\label{12D_expect_values}
\end{table}

\clearpage

\begin{table}[ht]
\caption{Energies (in cm$^{-1}$) of the $v=1,2$ HF stretching states of HF trimer, for intermolecular modes in the ground state, from 12D calculations using bases IA $(N_r = 10)$ and I $(N_r=8)$.}
\centering
\begin{tabular}{ccc}
\hline\hline
 IA & I & Assign.\\ \hline
3679.40 & 3679.04 & $\nu^{\rm HF}_{\rm sym}$  \\
3743.05 & 3742.69 & $\nu^{\rm HF}_{\rm asym}$  \\
7205.00 & 7201.27 & $2\nu^{\rm HF}_{\rm sym}$   \\ 
7225.31 & 7221.25 & $\nu^{\rm HF}_{\rm sym}+ \nu^{\rm HF}_{\rm asym}$ \\
 \{7400.04\}$^b$ & \{7397.49\}$^b$ & $2\nu^{\rm HF}_{\rm asym}(A_1')$ \\
7453.88 & 7452.90 & $2\nu^{\rm HF}_{\rm asym}(E')$ \\
\hline
\hline
\multicolumn{3}{l}{$^a$ The listed energies are all relative to the relevant ground-state energy.}\\
\multicolumn{3}{l}{$^b$ Significant basis-state mixing leading to small BSN.}\\

\end{tabular}
\label{tab_12D_intra_2}
\end{table}

\clearpage

\begin{table}[ht]
\caption{Intermolecular excitation energies (in cm$^{-1}$) in the ground-state (g.s.)  and $v=1$ $\nu^{\rm HF}_{\rm sym}$ and $\nu^{\rm HF}_{\rm asym}$ intramolecular manifolds of HF trimer, from 12D calculations using basis set IA.}
\centering
\begin{tabular}{cccc}
\hline\hline
Excitation & g.s. & $\nu^{{\rm HF}~a}_{\rm sym}$ & $\nu^{{\rm HF}~b}_{\rm asym}$ \\ \hline
$\nu_{as}$ & 170.91 & 183.82 & (179.28,185.75,180.76)$^c$\\
$\nu_{ss}$ & 186.89 & 198.94 & 196.35 \\
$\nu_{iab}$ & 501.43 & 536.30 & (526.67,534.31,529.43) \\
$\nu_{isb}$ & 776.70 &  \{827\}$^d$&  ---\\
$\nu_{oab}$ &  416.77 & 443.29  & (433.88,440.72,438.05) \\
$\nu_{osb}$ &  617.99 & 651.59 & 644.57 \\

  \hline
\hline
\multicolumn{4}{l}{$^a$ Energies relative to the intramolecular excitation energy of 3679.40 cm$^{-1}$.}\\ 
\multicolumn{4}{l}{$^b$ Energies relative to the intramolecular excitation energy of 3743.05 cm$^{-1}$.}\\ 
\multicolumn{4}{l}{$^c$ Combinations of two $E$-type excitations produce three distinct energy levels.}\\
\multicolumn{4}{l}{$^d$ Approximate value due to significant state mixing.}\\

\end{tabular}
\label{tab_inter_intra}
\end{table}

\clearpage

\begin{table}[ht]
\caption{Comparison of the fundamental frequencies (in cm$^{-1}$) of HF trimer from 12D calculations and spectroscopic measurements in the gas phase and neon (Ne) and argon (Ar) matrices. The calculated frequencies of the intermolecular modes, shown in the first six rows of the table, are for HF monomers in their ground intramolecular vibrational state.}
\centering
\begin{tabular}{lllll}
\hline\hline
Mode$^a$& calc.$^b$ & gas & Ne matrix$^c$  & Ar matrix$^d$ \\ \hline
$\nu_{as}$ ($\nu_7$) & 170.91 & &167 & 152.5\\
$\nu_{ss}$ ($\nu_3$)& 186.90 & & \\
$\nu_{iab}$ ($\nu_6$) & 501.44 & 495$^e$ & 477 & 446 \\
$\nu_{isb}$ ($\nu_2$) & 776.71 & &\\
$\nu_{oab}$ ($\nu_8$)&  416.77 &  &\\
$\nu_{osb}$ ($\nu_4$) &  617.99 &  602$^e$ & 590 & 560 \\
$\nu^{\rm HF}_{\rm sym} $ ($\nu_1$) & 3679.40 &  &\\
$\nu^{\rm HF}_{\rm asym}$  ($\nu_5$)  & 3743.05 & 3712$^f$&  3706 & 3702 \\

  \hline
\hline
\multicolumn{5}{l}{$^a$ The literature mode notation is given in parentheses.}\\ 
\multicolumn{5}{l}{$^b$ 12D calculations in this work, basis IA.}\\ 
\multicolumn{5}{l}{$^c$ Reference \onlinecite{ANDREWS92}.}\\ 
\multicolumn{5}{l}{$^d$ Reference \onlinecite{ANDREWS99}.}\\
\multicolumn{5}{l}{$^e$ Reference \onlinecite{ROYP14}.}\\
\multicolumn{5}{l}{$^f$ Reference \onlinecite{LISY86}.}\\

\end{tabular}
\label{tab_calc_obs}
\end{table}

\clearpage

\end{document}